%% file: paper.tex
\documentclass[conference]{IEEEtran}
\usepackage[table,xcdraw]{xcolor}
\usepackage{cite}
\usepackage{amsmath,amssymb,amsfonts}
\usepackage[linesnumbered,ruled,vlined]{algorithm2e}
\usepackage{textcomp}
\usepackage{tabularx}
\usepackage{lscape}

\usepackage{mathtools}
\usepackage{xspace}
\usepackage{xargs}
\usepackage{ifthen}
\usepackage{xifthen}
\usepackage{comment}
\usepackage{underscore}
\usepackage{subcaption}
\usepackage{listings}
\usepackage{float}
\usepackage{placeins}
\usepackage{afterpage}

\usepackage{etoolbox}
\usepackage[textsize=scriptsize,textwidth=1.1cm]{todonotes}
\usepackage[hidelinks]{hyperref}
\usepackage{multirow} 
\usepackage{circledsteps}

\def\BibTeX{{\rm B\kern-.05em{\sc i\kern-.025em b}\kern-.08em
    T\kern-.1667em\lower.7ex\hbox{E}\kern-.125emX}}

\usepackage[switch]{lineno}

\newif\ifversionCamera
\versionCamerafalse   

\begin{document}

\title{ETM$^{2}$: Empowering Traditional Memory \\ Bandwidth Regulation using ETM
}%

\iftrue
\author{%
	\IEEEauthorblockN{%
		Alexander Zuepke,
		Ashutosh Pradhan,
		Daniele Ottaviano,
		Andrea Bastoni,
		Marco Caccamo
	}
	\IEEEauthorblockA{%
		Technical University of Munich
	}%
    \IEEEauthorblockA{Email: \{alex.zuepke, ashutosh.pradhan, daniele.ottaviano, andrea.bastoni, mcaccamo\}@tum.de}%
}%
\else
\author{%
	\IEEEauthorblockN{%
		Omitted for blind Review
    }
}%
\fi

\maketitle


\setlength{\belowdisplayskip}{0.4em} 
\setlength{\abovedisplayskip}{0.4em} 

\input{0-notation}
\input{0-abstract}
\input{1-introduction}
\input{2-related}
\input{3-design}
\input{4-implementation}

\input{5-evaluation}

\input{6-conclusion}

\iftrue
\section*{Acknowledgments}
Marco Caccamo was supported by an Alexander von Humboldt Professorship
endowed by the German Federal Ministry of Education and Research.
\fi

\IEEEtriggeratref{37}
\bibliographystyle{IEEEtran}
\bibliography{paper}

\ifversionCamera
\else
\iftrue 
\input{7-appendix}
\fi 
\fi
\end{document}

%% file: 0-notation.tex
\definecolorseries{test}{rgb}{grad}[rgb]{.95,.55,.55}{11,11,17}
\resetcolorseries[10]{test}
\newcommand{\addtodoeditor}[1]{%
    \colorlet{#1}{test!!+!50}
    \expandafter\newcommand\csname#1\endcsname [1]{%
        \todo[color=#1,size=\tiny]{\sffamily\textbf{\uppercase{#1}:}
    ##1}\xspace%
    }
    \expandafter\newcommand\csname#1i\endcsname [1]{%
        \todo[inline, color=#1]{\sffamily\textbf{\uppercase{#1}:} ##1}\xspace%
    }
}


\addtodoeditor{az}
\addtodoeditor{ab}
\addtodoeditor{ot}
\addtodoeditor{ap}

\newcommand{\code}[1]{\mbox{\small\tt #1}\xspace}%
\newcommand{\bigo}[1]{\mbox{$\mathcal{O}(#1$)}\xspace}%
\newcommand{\bigolog}[1]{\bigo{\log{}#1}}%
\newcommand{\etal}{\mbox{\emph{et~al.\@}}\xspace}%
\newcommand{\wrt}{\mbox{\emph{w.r.t.\@}}\xspace}%
\newcommand{\eg}{\mbox{\emph{e.g.,\@}\@}\xspace}%
\newcommand{\ie}{\mbox{\emph{i.e.,\@}\@}\xspace}%
\newcommand{\cf}{\mbox{\emph{c.f.\@}}\xspace}%
\newcommand{\etc}{\mbox{\emph{etc.\@}}\xspace}%




\DeclarePairedDelimiter\ceil{\lceil}{\rceil}
\DeclarePairedDelimiter\floor{\lfloor}{\rfloor}

\newcommand{\optarg}[1][]{%
  \ifthenelse{\isempty{#1}}%
             {}
             {_{#1}}
}

\newcommand{\optpar}[1][]{%
  \ifthenelse{\isempty{#1}}%
             {}
             {(#1)}
}

\newtheorem{observation}{Obs.\@\xspace}

\newcommand{\FIXME}[1]{  \textbf{(FIXME: #1)}}
\newcommand{\memguard}{MemGuard\xspace}
\newcommand{\mempol}{MemPol\xspace}
\newcommand{\memcore}{MemCoRe\xspace}
\newcommand{\jailhouse}{Jailhouse\xspace}
\newcommand{\sdvbshort}{\mbox{SD-VBS}\xspace}
\newcommand{\sdvblong}{San Diego Vision Benchmark Suite\xspace}

\newcommand{\scupftchhit}{\code{SCU\_PFTCH\_CPU\_HIT}}
\newcommand{\scupftchmiss}{\code{SCU\_PFTCH\_CPU\_MISS}}
\newcommand{\scupftchacc}{\code{SCU\_PFTCH\_CPU\_ACCESS}}
\newcommand{\llcdcache}{\code{L3D\_CACHE}}
\newcommand{\llcdaccess}{\code{L3D\_ACCESS}}
\newcommand{\busaccessdsu}{\code{BUS\_ACCESS}}
\newcommand{\loneinner}{\code{L1D\_CACHE\_REFILL\_INNER}}
\newcommand{\loneouter}{\code{L1D\_CACHE\_REFILL\_OUTER}}
\newcommand{\cpuectrl}{\code{CPUECTRL}}
\newcommand{\lthreeews}{\code{L3D\_WS\_MODE}}
\newcommand{\lthreecachealloc}{\code{L3D\_CACHE\_ALLOCATE}}


\newcommand{\readInterf}{\textit{read}\xspace}
\newcommand{\writeInterf}{\textit{write}\xspace}
\newcommand{\modifyInterf}{\textit{modify}\xspace}
\newcommand{\prefetchInterf}{\textit{prefetch}\xspace}

\newcommand{\localization}{\textit{localization}\xspace}
\newcommand{\sift}{\textit{sift}\xspace}
\newcommand{\tracking}{\textit{tracking}\xspace}
\newcommand{\mser}{\textit{mser}\xspace}
\newcommand{\disparity}{\textit{disparity}\xspace}
\newcommand{\stitch}{\textit{stitch}\xspace}

\newcommand{\vga}{\textit{vga}\xspace}
\newcommand{\cif}{\textit{cif}\xspace}

\newcommand{\etmReg}{ETM\textsuperscript{2}\xspace}

\newcommand{\memguardBased}{PR\xspace} 
\newcommand{\mempolBased}{TB\xspace} 

%% file: 0-abstract.tex
\begin{abstract}
The Embedded Trace Macrocell (ETM) is a standard component of Arm's
CoreSight architecture, present in a wide range of platforms and
primarily designed for tracing and debugging.
In this work, we demonstrate that it can be repurposed to implement a novel
hardware-assisted memory bandwidth regulator,
providing a portable and effective solution to mitigate memory interference
in real-time multicore systems.

\etmReg requires minimal software intervention
and bridges the gap between the
fine-grained microsecond resolution of \mempol and
the portability and reaction time of interrupt-based solutions, such as \memguard.
We assess the effectiveness and portability of our design with an evaluation
on a large number of 64-bit Arm boards,
and we compare \etmReg with
previous works using a setup based on the San Diego Vision Benchmark Suite on the AMD Zynq UltraScale+.

Our results show the scalability of the approach and highlight the
design trade-offs it enables.
\etmReg is effective in enforcing
per-core memory bandwidth regulation and unlocks new regulation options
that were infeasible under \memguard and \mempol.
\end{abstract}
\vspace{-0.5em}
\begin{IEEEkeywords}
real-time system,
multicore,
memory bandwidth regulation
\end{IEEEkeywords}

%% file: 1-introduction.tex
\vspace{-0.5em}
\section{Introduction}
\label{sec:intro}

The problem of memory contention under real-time constraints has become
increasingly critical on modern embedded platforms,
which rely on shared memory hierarchies to achieve high performance
and energy efficiency.

Despite the increasing availability of architecture-defined
Quality of Service (QoS) mechanisms such as
Arm's Memory Partitioning and Monitoring (MPAM)~\cite{arm-mpam},
these features are not primarily designed for real-time workloads~\cite{ZiniCB:23}
and still exhibit many implementation-defined and
platform-specific characteristics~\cite{SBMYK:22},
which limit their predictability and portability across
different systems-on-chips (SoCs). 

In contrast, memory bandwidth regulation implemented in software
has been actively researched by the real-time systems community
over the past decade.
These techniques only rely on timers, interrupts, and the
Performance Monitoring Unit (PMU),
and are applicable to a wide range of platforms.

\memguard~\cite{YYPCS:13} first implemented an approach leveraging
Performance Monitoring Counters (PMCs) to estimate
memory activity and throttle cores using an interrupt-based approach.
This design allows \memguard-like techniques to promptly react
upon reaching per-PMC memory bandwidth thresholds,
but enforcing small bandwidth values and short regulation periods
incur overheads~\cite{YYPCS:16, memguard-ubicomm:16, SDZRRPHMGS:22, MemPol}.

\mempol~\cite{MemPol} 
removes the burden of interrupt management from the regulated cores and
achieved microsecond-scale granularity 
by moving
the regulation logic to an external processing element that periodically polls
the PMCs. Despite the much higher resolution and the possibility of monitoring
multiple PMCs at the same time, \mempol's polling-based
approach cannot immediately react when bandwidth thresholds are reached,
leading to \emph{overshooting}.
The recently proposed \memcore~\cite{IzhbirdeevH0ZHC24} extended
\mempol's approach by moving the control logic to an FPGA fabric
and exploiting cache-coherence events to achieve nanosecond-scale regulation.
Yet, this precision depends on platform-specific FPGA integration,
limiting its portability.

Contrary to these approaches, in this work
we show that to achieve \emph{both} fine-grained regulation resolution
and prompt reaction times,
additional computational resources external to the cores are not required.
This allows us to apply our strategy to a large number of Arm-based platforms.



\input{fig-intro_etm2}

We observe that the debug and trace infrastructure
on 64-bit Arm processors already provides (in hardware) all the
building blocks required to implement the basic components of
software-based memory bandwidth regulation techniques.
Specifically:
{\small\Circled{1}} integration with architectural elements such as PMCs,
{\small\Circled{2}} a control logic, and
{\small\Circled{3}} throttling mechanisms.

As illustrated in Fig.~\ref{fig:intro-regulation},
\memguard relies on the PMU for bandwidth monitoring and regulation
interrupts {\small\Circled{1}}.
The control logic is implemented by the Processing Elements (PEs)
using two interrupt handlers {\small\Circled{2}}, {\small\Circled{3}}
to throttle cores on PMC overflows and handle budget replenishment.
In \mempol, the external core implements
the control logic by polling the PMU {\small\Circled{1}}, {\small\Circled{2}},
as well as the throttling strategy by using the debug infrastructure to stall
and restart the PE {\small\Circled{3}}.
%

We show that the Embedded Trace Macrocell (ETM)~\cite{arm-etm},
a standard element of the Arm CoreSight~\cite{arm-coresight}
architecture and available on almost all
commercial off-the-shelf (COTS) Arm platforms,
can be cleverly exploited to perform most of memory bandwidth regulation activities
with minimal software support.
Our approach leverages the ETM's access to low-level events in the PE
to monitor memory bandwidth
by tracking memory transactions (see {\small\Circled{2}} in
Fig.~\ref{fig:intro-regulation}).
It also uses the ETM state-machine capability for
the regulation logic, generating interrupts to throttle cores
when bandwidth thresholds are exceeded {\small\Circled{1}}.
The only software support required
is a minimal interrupt
handler {\small\Circled{3}} (at operating system or hypervisor level)
to acknowledge ETM interrupts and stall the core.
Notably, no control logic is implemented in the software handler.
We termed our approach \etmReg: \emph{Empowering Traditional Memory Bandwidth Regulation using ETM}.


\etmReg inherits the strengths of existing regulators.
Like \mempol, it achieves microsecond-scale precision and
can combine multiple monitoring dimensions (cache refills and write-backs).
Like \memguard, it is interrupt-based and free from overshooting caused 
by periodic polling of PMCs like in \mempol.
Despite lacking the fine-grained physical-address awareness of FPGA-based
approaches (\eg \memcore), it preserves full compatibility with commercial
Arm-based systems, requiring no hardware modification or FPGA integration.


In summary, the key contributions of this paper are:
\begin{itemize}
    \item ETM-based regulation: We leverage the CoreSight ETM to implement a hardware-assisted, memory bandwidth regulator directly within the existing debug infrastructure.
    \item Interrupt-based fine-grained regulation: We demonstrate microsecond-scale regulation intervals with an interrupt-driven approach that avoids PMC-polling overshooting.
    \item Two regulator designs, \emph{periodic replenishmen (\memguardBased)} (\emph{\`a la} \memguard) and \emph{token-bucket (\mempolBased )}  (\emph{\`a la} \mempol).
    \item Wide platforms coverage: \etmReg has been implemented on a broad range of 64-bit Arm SoCs, relying solely on standard CoreSight components.
\end{itemize}


The remainder of this paper is structured as follows.
Sec.~\ref{sec:related-work} reviews background on memory-bandwidth regulation, focusing on software-based approaches, and provides the necessary context on Arm CoreSight and the ETM.
Sec.~\ref{sec:design} presents the design of \etmReg, detailing how CoreSight components are repurposed for regulation.
Sec.~\ref{sec:implementation} describes our implementation across multiple 64-bit Arm platforms.
Sec.~\ref{sec:eval} evaluates \etmReg in terms of correctness, resolution limits, and its comparison against \memguard and \mempol, using both micro-benchmarks and SD-VBS workloads.
Finally, Sec.~\ref{sec:conclusion} concludes the paper and outlines directions for future work.

%% file: fig-intro_etm2.tex
\begin{figure}[t!]
    \centering
    \includegraphics[width=0.95\columnwidth]{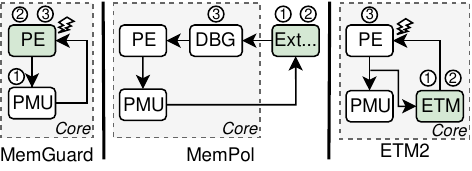}
    \caption{High-level comparison among traditional memory bandwidth regulation mechanisms \memguard and \mempol, and the presented \etmReg. High-lighted in green is the regulation logic. Lightning arrows indicate interrupts.}
    \label{fig:intro-regulation}
\vspace{-0.5em}
\end{figure}

%% file: 2-related.tex
\section{Background and Related Work}
\label{sec:related-work}

Memory bandwidth regulation has been extensively studied
to improve timing predictability and reduce interference, especially in
multiprocessor systems-on-chip (MPSoC) systems.

Most software-based approaches use PMCs to measure per-core memory accesses
and subsequently trigger regulation activities.
Starting with \memguard~\cite{YYPCS:13}
(discussed in detail in Sec.~\ref{sec:memguard}),
several other studies have experimented
with variants of the same technique in the context of hypervisors~\cite{memguard-ubicomm:16, xvisor-biondi:18, bao:20},
or to enact regulation based on reads or writes~\cite{BY:19}.

Compared to \memguard, the approach proposed
by \mempol~\cite{MemPol} (see Sec.~\ref{sec:mempol}) implements
a finer-grained regulation using a polling-based strategy.
PMU-based memory-regulation policies can only approximate
the effective utilization of the DRAM controller~\cite{STDM:20},
but very few platforms allow to precisely quantify
memory bandwidth using performance counters
co-located within the DRAM~\cite{SDZRRPHMGS:22}.

Orthogonally to PMC-monitoring,
bandwidth regulation has also been implemented using
the quality of service (QoS) infrastructure available on some
MPSoC~\cite{arm-qos400, Tegra-Regul:17, QoS-Cazorla:21, ZCCB:22, PradhanOJHZZBC25},
or architectural features such as Arm's MPAM~\cite{arm-mpam}
or Intel's RDT~\cite{intel-rdt}.
Nonetheless, the limited availability
of these mechanisms in embedded systems and their platform-specific,
often unpredictable behavior~\cite{SBMYK:22,MPAMWiP}, continue to hinder their adoption.

Although this work focuses on hardware-assisted, software-based
bandwidth regulation, several prior studies have proposed
dedicated FPGA- or hardware-extensions (\eg~\cite{MITTS:16, DBLP:conf/date/CardonaHAC19, BRU:20, wessman2021risc})
that implement regulation directly in hardware,
or that improve worst-case latency of contended memory transactions
with a dedicated memory-controller design (\eg \cite{DRAMbulism:20, MEDUSA:15}).
%
On PS-PL platforms (\eg~\cite{zcu-102}), following~\cite{HRSM:21},
\memcore~\cite{IzhbirdeevH0ZHC24} uses the FPGA to monitor
cache-coherency traffic and throttle cores similarly to \mempol.
In contrast,~\cite{Baryshnikov2018FPGA} only monitors cache-miss from FPGA using the CoreSight infrastructure.

\subsection{\memguard}
\label{sec:memguard}
\memguard ~\cite{YYPCS:13} enforces memory bandwidth regulation by assigning
each core a budget of memory transactions within a fixed regulation period.
The budget is expressed in terms of PMC events,
such as last-level cache refills or write-backs. 
The memory budget is consumed when cores perform memory transactions
and is replenished according to the regulation period.
\memguard relies on two interrupts: one is triggered when the PMU counter
reaches the assigned budget, upon which the core is temporarily throttled;
the other is a periodic timer interrupt that resets the PMU counter
and resumes the core if it was previously throttled.
This design provides per-core bandwidth control without
requiring hardware support beyond standard PMCs,
but using more than one counter requires individual budgets for each PMU-based metric.
As such, \memguard allows only limited combinations (\ie \emph{min}, depending on which PMC fires first)
of the contributions of different counters~\cite{BY:19}.
The interrupt-based mechanism adopted by \memguard
introduces an overhead that increases with smaller regulation periods (more timer interrupts),
or with smaller budget assignments (more PMU interrupts due to smaller leeway on bursts)~\cite{YYPCS:13, MemPol, SDZRRPHMGS:22}.
As a result, 
\memguard's periods are coarse-grained (typically 1~ms~\cite{MemPol}),
limiting the achievable temporal resolution and allowing
unregulated bursts of memory-intensive activity.
    
\subsection{\mempol}
\label{sec:mempol}
\mempol~\cite{MemPol,ZuepkeBCCM24} addresses the granularity limitations of
\memguard by decoupling monitoring and enforcement from the regulated cores. 
Instead of PMU-triggered interrupts, \mempol periodically
polls PMCs from an external control core
(\eg one of the real-time cores available on MPSoCs~\cite{zcu-102})
and uses debug interfaces (\eg Arm's CoreSight~\cite{arm-coresight}) to halt and resume
the target cores when their budget is exceeded.
\mempol guarantees a minimum bandwidth over time, using equivalently a
sliding window~\cite{MemPol} or a token-bucket~\cite{IzhbirdeevH0ZHC24}
regulation strategy.
The polling-based design enables regulation at microsecond-scale periods,
while avoiding any overhead on the regulated workloads (excluding the throttling if the budget is exceeded).
\mempol can also flexibly combine multiple PMCs at runtime to derive more
accurate estimates of memory pressure than single-counter schemes~\cite{MemPol, PradhanOJHZZBC25}.
However, due to its polling behavior,
\mempol cannot prevent \emph{overshooting} the target budget before
the regulation can be enacted.

\subsection{Regulation Granularity}

Memory bandwidth regulation at small scales (\eg microseconds, function calls)
allows for better integration
with hardware-based QoS schemes at interconnect level~\cite{arm-qos400}.
However,
regulation at small scales
is inherently more pessimistic
than regulation at coarse scales (\eg milliseconds, task scheduling):
due to the much smaller budgets,
a fine-grained regulation has less leeway on short bursts of memory accesses,
\eg \code{memcpy} or \code{memset} operations,
and throttles cores for a longer time.
\mempol favors small scales to reduce overshooting, 
but allows to compensate bursts with its sliding window or token-bucket mechanisms.
%
%
%
%
%
Thus the choice of regulation mechanism and granularity eventually depends on the timing requirements of a system, 
\eg fine-grained regulation can be beneficial for tasks with periods of 100~µs or below, or when co-regulating with hardware-based QoS.

\subsection{Arm CoreSight}

\input{fig-coresight_highlvl}

Arm defines \emph{CoreSight} for debugging of its cores
and specifies \emph{invasive} and \emph{non-invasive} debug~\cite{arm-coresight}.
Invasive debug gives a hardware debugger or a self-hosted software debugger full access to a core's internal state.
Non-invasive debug comprises performance monitoring and tracing.


Fig.~\ref{fig:coresight} compactly describes the CoreSight components of 64-bit Arm Cortex-A processors~\cite{arm-coresight}.
%
The debug interface (DBG) offers low-level access to the core's registers for debugging purposes.
The PMU provides six counters that can be programmed to count a larger number of PE-specific events and signal interrupts.
The ETM monitors control flow-specific information of the PE and has access to the same PE-specific events as the PMU.
This is used to generate trace events for other components outside of the core
that
%
collect the traces and make them available for further processing.
The Cross-Trigger Interface (CTI) 
provides access to low-level control signals to synchronize DBG, PMU, and ETM events among cores in multicore setups through the Cross-Trigger Matrix (CTM) 
and
raises interrupts to the Generic Interrupt Controller (GIC).
Two important interrupts are PMUIRQ to signal PMC overflows and CTIIRQ to forward selected events at the CTI.

In Cortex-A processor cores, DBG, PMU, ETM, and CTI are always part of the core
and run in the same clock domain and at the same speed as the PE.
CoreSight components are accessible through core internal registers (DBG and PMU only) or as memory-mapped devices. 
The memory-mapped interfaces are accessible through a low-bandwidth Debug Advanced Peripheral Bus (APB). 
Unless debugging is disabled or locked down for security purposes,
both hardware debuggers and the cores can access the CoreSight components.

Note that CoreSight access is typically disabled in production deployments,
as it allows to bypass security perimeters of the kernel or hypervisor~\cite{ning2021revisiting}.
System integrators must therefore carefully consider the risks of giving the cores access to non-invasive CoreSight components for their systems.\footnote{Access for hardware debuggers (\eg JTAG) is managed by TAP controllers that typically implement security gateway functionality in the SoC.}

\subsection{ETM Background}
\label{sec:etm-background}

The ETM monitors internal signals of the PE and a diverse set of resources to implement \emph{filters} and \emph{triggers} for tracing~\cite{arm-etm}.
%
Filters define \emph{``what''} to trace,
while triggers define  events to start and stop tracing (\emph{``when''}).
Additional registers configure frequency and level of detail of the generated trace data.
%
%
Fig.~\ref{fig:etm} shows the ETM resources relevant for this work,
which are the same on all 64-bit Cortex-A processors.

\textbf{External inputs:}
The ETM is connected to four external input signals from the CTI and a set of low-level signals from the PE that also feed the PMU.
The ETM can monitor up to four of these signals \emph{OR}-ed together.
Unfortunately,
on out-of-order cores such as Cortex-A72, A76 and A78,
events related to cache accesses have multiple low-level signals that can become active at the same time.

\textbf{Address comparators:}
Eight address comparators match configured addresses or can be used in pairs as address range comparators.
Cortex-A cores implement only matches on instruction fetches and for virtual addresses.

\textbf{Resource selectors:}
16 resource selectors function as multiplexer from specific \emph{groups} like external inputs, single or pairs of address comparators, counters, or sequencer states.
From the selected group, a selector can combine (logical \emph{OR}) up to 16 specific resources.
Selectors can be used in pairs with a Boolean operation, \eg AND or OR,
and can be configured to negate their outputs.
The first two selectors are hardwired to always provide \emph{TRUE} (1) and \emph{FALSE} (0) signals respectively.

\textbf{Counters:}
The ETM contains two 16-bit 
counters that count \emph{down} when its configured input signal (resource selector) is active.
The counter ``fires'' when it reaches the value zero.
Each counter has a programmable reload value and can be configured to self-reload or on a trigger signal from a resource selector.
Depending on whether a counter reloads itself, the counter outputs a pulse (self-reload mode) or level signal (not self-reloading).
%
Additionally, the two timers can be chained together into a single 32-bit counter.

\input{fig-etm_architecture}

\textbf{Sequencer:}
The ETM provides a state machine with four states, $0 \Leftrightarrow 1 \Leftrightarrow 2 \Leftrightarrow 3$.
The sequencer allows configuring resource selectors for the three forward and the three backward transitions, and to reset to state $0$.
The sequencer can change multiple states within the same cycle, as long as the conditions for state transitions evaluate to true. Forward movements take precedence over backward movements, and resets take precedence over both.
The sequencer stays in its state when no other conditions evaluate to \emph{TRUE} in the current cycle.
Sequencer states can be configured as inputs for resource selectors (\emph{OR}-ed together) with a level signal as output.

\textbf{External outputs:}
The ETM can drive four ETM output signals to the CTI
depending on the configured resource selectors.
The CTI can be configured to route these signals 
into the PE (to halt or resume the core for debugging purposes), 
to other cores' CTIs via the CTM,
to raise the CTIIRQ interrupt to the GIC,
or even back into the ETM.

\textbf{Programming model:}
The ETM is programmed by configuring all described resources,
using the \emph{FALSE} selector to disable unused resources.
In this work, we configure all trace resources, but we disable the generation
of trace data, 
as it is not
relevant for bandwidth regulation.
%
Programming the ETM can be particularly cumbersome
due to the mixed use of pulses and level outputs and edge- and level-triggered inputs.
%


%% file: fig-coresight_highlvl.tex
\begin{figure}[t!]
    \centering
    \includegraphics[width=0.95\columnwidth]{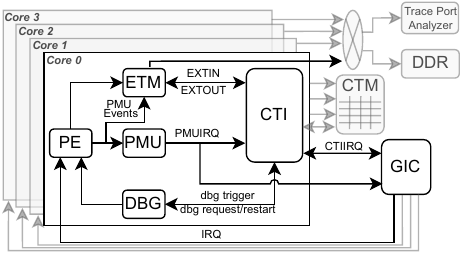}
    \caption{Arm CoreSight components, highlighting the modules used by \etmReg.}
    \label{fig:coresight}
\vspace{-.9em}
\end{figure}

%% file: fig-etm_architecture.tex
\begin{figure}[t!]
    \centering
    \includegraphics[width=0.95\columnwidth]{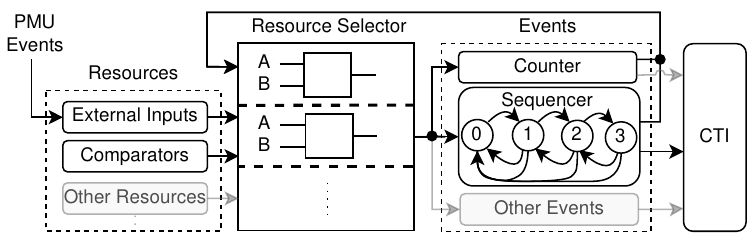}
    \caption{Internal ETM architecture, highlighting the modules used by \etmReg.}
    \label{fig:etm}
\vspace{-1.2em}
\end{figure}

%% file: 3-design.tex
\input{fig-state_machines}

\section{Design}
\label{sec:design}



The core logic of software-based memory bandwidth regulators like
\memguard and \mempol (see Fig.~\ref{fig:intro-regulation}) is
straightforward.
These regulators a) keep track of memory bandwidth usage monitoring last-level cache (LLC) activity,
and b) throttle cores (\eg via interrupt) that exceed their assigned budget
until they can resume normal operations (\eg the budget is \emph{replenished}).
These steps could be implemented in hardware
\emph{near} the core where the required data is already available.

While researching the ETM's capabilities for bandwidth regulation,
we observed that the ETM provides the minimum amount of resources
to implement such a regulator,
\ie configurable access to PMU data,
two counters to account for LLC activity and time,
and two different mechanisms to halt the core,
debug halt and CTIIRQ.
The driving research questions then became
where the limitations and advantages of the ETM are,
and
which types of regulator designs are possible.

\subsection{External Inputs to the ETM}
\label{sec:design:inputs}
Like the PMU,
the ETM has access to the same low-level signals in the core regarding last-level cache activity.
But unlike the PMU,
the ETM is limited to monitor four low-level input signals
that are logically \emph{OR}-ed together
rather than the PMU's six counters and proper adders.\footnote{
    For an event comprising $N$ signals, the PMU employs a combinatorial network and adds the resulting value in the range $0$ to $2^{N-1}$ to the PMC.
}

On first generations of Arm cores such as Cortex-A53, A57, and A72,
memory bandwidth regulation
relies on two PMU events,
\ie L2 cache refills and L2 cache write-backs,
that are available as two input signals to the ETM~\cite{arm-a53trm, arm-a57trm, arm-a72trm}.
However,
newer generations, such as Cortex-A76 and A78, 
require different PMU events~\cite{PradhanOJHZZBC25}
that expose three or more low-level signals to drive the PMU counters~\cite{arm-a76trm, arm-a78trm}.
This raises the important question of whether
accounting
\emph{OR}-ed input signals is a sufficient metric for regulation.
We will demonstrate feasibility and discuss this in detail in Sec.~\ref{eval:correctness}.

Furthermore, the monitoring of LLC activity available to the ETM
cannot be leveraged to realize a global regulation policy like in \mempol.
Given a typical four-core cluster with four input and four output signals,
propagating each core's individual throttling state
through the CTM to the other cores in the cluster
requires three of the four inputs and
would reduce the LLC activity information to a single input signal.

\subsection{Counters}
The ETM provides two 16-bit counters (Sec.~\ref{sec:etm-background})
that we use to count PMU events (Sec.~\ref{sec:design:inputs})
and time using the \emph{TRUE} input selector.
We program both timers in self-reloading mode 
with the core's \emph{budget} (number of cache lines) 
and \emph{period} for budget replenishment (CPU cycles).
Two limitations become obvious:
(i)
16-bit counters provide an upper bound for the duration of the period,
\eg 32.7~µs for a 2000 MHz core,
and (ii) self-reloading may cause overflows for small budget values
when the counter keeps incrementing after the core's budget was exceeded,
\eg due to overheads of an interrupt handler
or flushing of internal queues in the core's memory subsystem.
We look at the latter problem in detail
in Sec.~\ref{eval:boundaries}.

\subsection{External Outputs of the ETM}
\label{sec:design-extout}
The ETM's external output can drive the low-level input pins to stop (EDBGRQ) or resume (DBGRESTART)
the core for debug purposes, or raise an interrupt (CTIIRQ).

A fast low-level stop/resume mechanism would be favorable for memory bandwidth regulation,
however a debug halt requires an explicit acknowledgment signal (DBGACK) before the core can be resumed~\cite{MemPol},
and this signal is not connected to the CTI.
Instead, the debug halt state must be acknowledged by a write to a 32-bit register
in the CTI's memory-mapped region.
This would require busmaster capabilities not available to the ETM,
and this obviously cannot be combined with an interrupt handler on the core to write to the acknowledgement register either,
as the core is already halted.
We therefore did not follow this approach,
but opted instead for triggering the CTIIRQ interrupt
to throttle the core in the interrupt handler.

Unfortunately, the routing of the CTIIRQ signal to the GIC can be problematic:
on some platforms (\eg the Nvidia Orin) the CTIIRQ is not available, while on
others (\eg the NXP i.MX8 and i.MX93) it is shared among the cores.
In this paper, we did not further attempt to find a solution for these systems.

\subsection{Sequencer}
When ETM counters reach zero, they provide a pulsed output signal before reloading on the input event.
Such a pulsed output is not observable by an interrupt handler.
Therefore, we have to use sequencer states to indicate that the core is within or exceeds its configured budget.
With two states, we can design a MemGuard-like regulation
which we present in Sec.~\ref{sec:design-memguard}.
However, as the sequencer has four states,
we evaluated the use of the additional states for regulation purposes, and also designed a MemPol-like regulation
which we present in Sec.~\ref{sec:design-mempol}.

Sequencer states must also be used 
for address-based filtering when using the address range comparators,
\eg to disable regulation in specific memory regions, or in the OS kernel or hypervisor.
The address range comparators emit pulses at different times than PMU input signals,
thus the two signals cannot be combined by \emph{OR}-ing them together as a counter input.
We tackle these challenges with a design
to enable accounting and regulation only in user space
(Sec.~\ref{sec:design-address-based}).


\subsection{Periodic Replenishment (\memguardBased) Regulation}
\label{sec:design-memguard}
The \etmReg \memguardBased regulator
provides \emph{periodic replenishment} of the core's budget.
Fig.~\ref{fig:design-memguard} sketches the design.
The two counters drive the sequencer state machine in either forward (budget exhausted)
or backward direction (budget replenished).
We use just two states, 0 and 3, 
and raise CTIIRQ in state 3.
The interrupt handler in the OS or hypervisors polls the sequencer state status register and waits for the state to change back to zero before returning.
Important in this design is that the replenishment also resets the budget counter, but we do this only when the sequencer is in state 0.
This allows us to accumulate any overuse of bandwidth while in state 3,
as the PMU counter resets on the transition $0 \Rightarrow 3$,
and reduce the budget from the next regulation period accordingly.

The general behavior of this regulator mimics \memguard's behavior,
except that we don't require interrupts to replenish the budgets,
as the ETM-based design implements the replenishment directly in hardware.
This solves one of the major performance issues when \memguard
is used with a small regulation period,
as there is no overhead for replenishment for the core under regulation.
Compared to \memguard,
a limitation of the presented regulator is that the length of the period is limited by the 16-bit counter. 


\subsection{Token Bucket (\mempolBased) Regulation}
\label{sec:design-mempol}
The \etmReg token bucket (\mempolBased) regulator
uses the additional two sequencer states
to model a \emph{token bucket} regulation.
Such a design allows the application to compensate for memory access bursts
over multiple regulation periods.
%
In this model, the bucket's fill counter can be interpreted
as a rational number, where the sequencer state represents
the integer part and the counter value represents the fractional part,
and replenishment occurs in integer steps,
\eg from 2.427 to 1.427.
The two counters drive the state machine forwards resp.\@ backwards.
An important difference to the \memguardBased regulator 
is that \mempolBased does not reset the budget counter, 
as this would break the logical bucket model
(the replenishment counter could reset the budget, for example, from 2.427 to 1.0).

Fig.~\ref{fig:design-mempol} shows the design for three variants
of the \mempolBased regulator.
\mempolBased~3:1 uses three states to model the acceptable range
and one state to indicate that the budget is exceeded.
\mempolBased~2:2 uses two states for each range.
\mempolBased~1:3 uses a single state for the acceptable range
and three states for budget overuse.
The motivation for the three different designs
is their behavior \wrt counter overflows at very small budgets or very large budgets,
which we will discuss further in Sec.~\ref{eval:boundaries}.


\subsection{Address-based Regulation (PR-user)}
\label{sec:design-address-based}
We also evaluated the feasibility of address-based regulation,
\ie providing fine-grained control for accounting and/or replenishment
when the core executes in specific memory regions.
Such an approach can for example distinguish between execution in user mode or kernel mode resp.\@ hypervisor mode~\cite{IzhbirdeevH0ZHC24}.
The address range comparators in the ETM provide filtering of instruction execution
based on virtual address ranges,
the core's current execution state (\ie EL0 to EL3),
and the currently scheduled task or VM IDs based on internal registers
(\emph{CONTEXTIDR} for task IDs and the \emph{VMID} portion in \emph{VTTBR} for VM IDs).
However,
address range comparators only activate (pulse) on instruction execution,
therefore we need sequencer states to get a stable level output
to combine this information with the pulses from PMU events or counters
in resource selector pairs.

For accounting of memory accesses only in user space,
dubbed \memguardBased-user,
we configured the address range comparators
to follow the split in the virtual address space between user and kernel space,
and use the following four states: 
$0$: $<$budget/kernel;
$1$: $<$budget/user;
$2$: $\ge$budget/user;
$3$: $\ge$budget/kernel.
%
Transitions $1 \Leftrightarrow 2$ follow the \memguardBased regulator design.
Transitions $0 \Leftrightarrow 1$ resp.\@ $2 \Leftrightarrow 3$
happen when the core changes execution modes.
Budget replenishment additionally drives the state machine $(2 \lor 3) \Rightarrow 0$, 
and the sequencer adjusts to $1$ on the next instruction fetch from user space.
%

%% file: fig-state_machines.tex
%
%
%
\begin{figure*}[t!]%
\begin{minipage}{.49\textwidth}%
    \centering%
    \includegraphics[width=0.95\columnwidth]{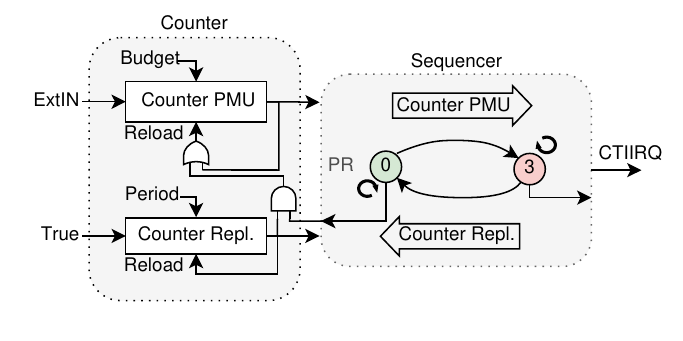}%
    \caption{\etmReg \memguardBased regulation:
    The design follows MemGuard
    by using two counters for budget accounting and periodic replenishment.
    The sequencer turns the counter transitions (edge)
    into stable output signals (level).
    State 3 indicates throttling to the OS or Hypervisor via CTIIRQ.
    }%
    \label{fig:design-memguard}%
\end{minipage}%
%
\hfill%
%
\begin{minipage}{.49\textwidth}%
    \centering%
    \includegraphics[width=0.95\columnwidth]{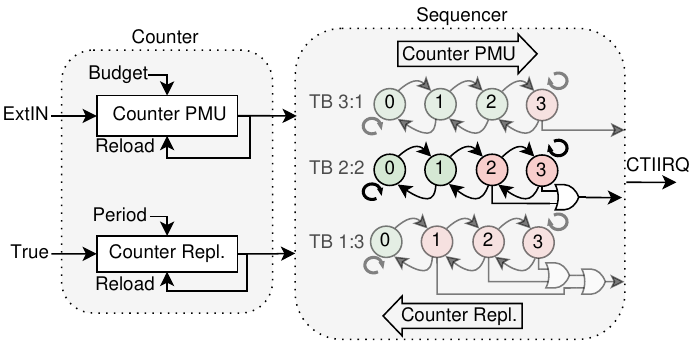}%
    \caption{\etmReg \mempolBased regulation:
    The design follows MemPol's token-bucket approach
    with the two counters driving the state machine transitions back and forth.
    Depending on the \mempolBased variant,
    states 1 to 3 indicate throttling.
    \etmReg TB 2:2 provides robustness for both small and large regulation budgets.
    }%
    \label{fig:design-mempol}%
\end{minipage}%
\vspace{-1.1em}%
\end{figure*}%

%% file: 4-implementation.tex
\section{Implementation}
\label{sec:implementation}
\etmReg needs access to the memory-mapped CoreSight registers
from the OS or hypervisor.
Consequently, the security implications identified in \mempol~\cite{MemPol}
also apply: the CoreSight infrastructure can bypass TrustZone protections,
and system integrators must account for this.

Each core needs a dedicated CTIIRQ line at the GIC (see Fig.~\ref{fig:coresight}).
%
This interrupt
must be configured as level-triggered on Cortex-A53 to A72
and edge-triggered for newer 
cores such as Cortex-A55 or A76.
In older SoCs, the interrupt is often connected as shared peripheral interrupt (SPI).
Newer SoCs follow Arm's recommendation and provide CTIIRQ as private peripheral interrupt (PPI)~24 to the GIC.

We observed that the ETM stops counting when the core enters an idle state by executing \emph{wait for interrupt (WFI)} or \emph{wait for event (WFE)} instructions.
In idle state, the core is not causing interference, but also the automatic replenishment stops.
The kernel or hypervisor should therefore
enter idle state only during replenished state,
which is likely the case when the core has no other pending activity and pending interrupts.

Lastly, as the regulator derives its internal timing for budget replenishment
from the \emph{TRUE} resource selector, and therefore indirectly from the core's current clock speed,
the core should be configured to a fixed frequency.

\subsection{Software Components}
The \etmReg implementation consists of two software components:
(i) a component to set up CoreSight devices and program the core's regulation values,
and
(ii) an interrupt handler for CTIIRQ that polls the ETM sequencer until the ETM leaves the regulation-specific throttling state. It afterwards acknowledges the CTIIRQ by writing to the CTI interrupt acknowledge register \emph{CTIINTACK} and in the GIC.

Our reference implementation\footnote{
\url{https://github.com/rtsl-cps-tum/rtas2026-etm2}}
consists of
an application to program the ETM from user space
and
a Linux kernel module that implements the interrupt handler.
The application opens \emph{/dev/mem} and configures the memory-mapped CoreSight devices for the desired regulator design and the cores' bandwidth settings.
The kernel module is configured through board-specific settings for CTIIRQ in the DTB (interrupt number and type, and core affinity) and regulation-specific settings to indicate throttling states.

\subsection{Programming of DBG, PMU, ETM, and CTI}
\label{sec:impl-programming}
CoreSight devices are programmed by first unlocking
the DBG, CTI, ETM, and PMU devices
via their memory-mapped interfaces
and programming the devices into an initial state
with disabled functionality
(see Arm specification~\cite{arm-coresight}).
%

We program the CTI to connect the ETM external outputs 1, 2, and 3 to the CTIIRQ input signal of the CTI.
We leave all other signals of the core in a disconnected state.

For the ETM, we program the external inputs to monitor core-specific signals (see Sec.~\ref{sec:impl-core-settings}),
the address range comparators (if needed),
the resource selectors,
the counters,
and
the sequencer state transitions.
We connect ETM outputs to the respective sequencer states that indicate throttled state,
and finally enable the ETM.

For the ETM to observe PMU events and thus the regulation to work,
the PMU must also be initially programmed to enable the export of PMU events to the ETM
by setting the \emph{X}~bit (export) in the PMU's PMCR control register.
Afterwards,
the OS or hypervisor can configure and use the PMU independently from the ETM,
as they are separate hardware units.

The \etmReg \memguardBased regulator requires six resource selectors,
with free resources to implement address-based filtering.
The \mempolBased regulator uses one resource selector to configure the external inputs for the PMU counter and
six pairs of resource selectors
to define the signals that drive the sequencer state machine back and forth.
Extensions to the \mempolBased design are not possible, as there are no more ETM resources available.

\subsection{Core-Specific Regulation Settings}
\label{sec:impl-core-settings}

For memory bandwidth regulation,
we consider four memory access operations:
%
\emph{Read} refers to cache refills from load operations or instruction fetches.
\emph{Prefetch} refers to cache refills from prefetching
and allow stressing the read capabilities of the memory subsystem:
the core does not have to wait for the result 
and handles prefetches in an out-of-order manner,
even on in-order cores.
Arm cores do not fetch a cache line from memory
when the full content of a cache line is overwritten,
\eg in \code{memset}.
Instead, the core allocates a cache line
that is eventually written back.
We denote this type of access as \emph{write} operation.
Lastly, \emph{modify} describes a partial write to a cache line, requiring the cache line to be fetched from memory and eventually be written back, causing twice the memory activity.

For early generations of 64-bit Arm cores,
namely Cortex-A53, A57 and A72,
which have private L1 instruction and data caches
and use a shared L2 cache as LLC,
we monitor the core's L2 activity as
\[
B_{A53}=B_{A72}=\emph{L2D_CACHE_REFILL} + \emph{L2D_CACHE_WB}.
\]
This is in line with previous work (\eg \cite{MemPol, IzhbirdeevH0ZHC24})
and was shown to accurately track a core's memory activity.

\textbf{Cortex-A53:}
The PMU events are provided as external PMU inputs 21 and 22 to the ETM.

\textbf{Cortex-A57 and A72:}
Cortex-A57 and A72 cores have the same programming model for PMU and ETM.
The events are provided on ETM PMU inputs 24 and 25.

For newer \emph{Dynamic Shared Unit} (DSU)-based cores,
namely Cortex-A55, A76 and A78,
Arm extended the memory subsystem to three cache levels
with private split L1 caches, private unified L2 caches,
and a shared L3 cache as LLC.
As no per-core PMU event tracks the LLC activity accurately,
we use the same PMU settings for bandwidth monitoring as in~\cite{PradhanOJHZZBC25}.
Note that having a private L2 cache is optional for Cortex-A55 cores.
Our regulation scheme remains valid regardless of
the presence of an L2 cache,
as the DSU cache always uses the position of the L3 cache in the programming model~\cite{arm-a55trm}.



\input{tab-boards_table}

\input{tab-ETM_correctness_table}


\textbf{Cortex-A55:}
For this core, we use three models:
\[
B_{A55}^{\emph{pessimistic}} = \text{2} \times \emph{L3D_CACHE_ALLOC},
\]
\[
B_{A55}^{\emph{moderate1}} = \frac{1}{4} \times \emph{BUS_ACCESS},
\]
\[
B_{A55}^\emph{moderate2} = \emph{L3D_CACHE_ALLOC} + \emph{L3D_CACHE_REFILL}.
\]
ETM PMU inputs are 33 $(B_{A55}^{\emph{pessimistic}})$, 23 $(B_{A55}^{\emph{moderate1}})$, resp.\@ 33 and 34 $(B_{A55}^{\emph{moderate2}})$~\cite{arm-a55trm}.
The models progressively decrease the overestimation of \emph{writes} and \emph{reads}~\cite{PradhanOJHZZBC25}. 

\textbf{Cortex-A76 and A76AE:}
Here we use:
\[
B_{A76}^{\emph{pessimistic}} = \text{2} \times \emph{L2D_CACHE_WR},
\]
\[
B_{A76}^{\emph{moderate1}} = \emph{L2D_CACHE_WR} + \emph{L3D_CACHE_REFILL},
\]
\[
B_{A76}^\emph{moderate2} = \emph{L2D_CACHE_WR} + \emph{L3D_CACHE_ALLOC}.
\]
Similarly to the A55, these models are progressively more accurate
in estimating \emph{writes} and \emph{reads}~\cite{PradhanOJHZZBC25}.
%
%
ETM PMU inputs for \emph{L2D_CACHE_WR} comprises two signals 73 and 74,
for \emph{L3D_CACHE_REFILL} also comprises two signals 158 and 159,
and \emph{L3D_CACHE_ALLOC} uses one signal 157~\cite{arm-a76trm}.

We note that MemGuard is restricted to using the \emph{pessimistic} and \emph{moderate1} settings on Cortex-A55 and the \emph{pessimistic} setting on Cortex-A76 due to the combination of PMUs required by the other settings.

\textbf{Cortex-A78 and A78AE:}
On this core type, we use
\[
B_{A78}^\emph{pessimistic} = 2\times\emph{L2D_CACHE_WR}.
\]
The model regulates \emph{modifies} accurately,
but overestimates \emph{reads} and \emph{writes} by 100\%~\cite{PradhanOJHZZBC25}.
%
ETM PMU inputs are 103, 104, and 105~\cite{arm-a78trm}.
We cannot use better models from~\cite{PradhanOJHZZBC25}, as
\[
    B_{A78}^\emph{moderate1} = \emph{L2D_CACHE_WR} + \emph{L3D_CACHE_REFILL},
\]
requires monitoring of six ETM PMU inputs.
Likewise,
\[
B_{A78}^\emph{moderate2}=\frac{1}{4}\times\emph{BUS_ACCESS_WR} + \emph{L3D_CACHE_REFILL}
\]
cannot be used due to the fractional factor in the sum.

%% file: tab-boards_table.tex
%
%
\begin{table*}
\centering
\caption{Characterization of the set of platforms used to test \etmReg. 
}
\begin{tabular}{lcccccc}
\hline
Board & Full Name & Cores & Speed [MHz] & CTIIRQ & IRQ Latency [cycles] & Status \\
\hline
ZCU102 & AMD Zynq UltraScale+ & 4xA53 & 1200 & SPI & 81 & \textcolor{green!60!black}{Works} \\
i.MX8M & Coral AI dev board & 4xA53 & 1200 & Shared SPI & - & Not supported \\
i.MX8MP & Debix Model A & 4xA53 & 1200 & Shared SPI & - & Not supported \\
i.MX93 & Debix Model C & 2xA55 & 1692 & Shared SPI & - & Not supported \\
S32G2 & MicroSys S32G274AR2SBC2 & 2x (2xA53) & 1000 & PPI & 80 & \textcolor{green!60!black}{Works} \\
LX2160A & Microsys miriac SBC-LX2160A & 8x (2xA72) & 2000 & PPI & 259 & \textcolor{green!60!black}{Works} \\
TI AM62x & BeaglePlay & 4xA53 & 1250 & PPI & 131 & \textcolor{green!60!black}{Works} \\
TI AM67x & BeagleY-AI & 4xA53 & 1250 & PPI & 162 & \textcolor{green!60!black}{Works} \\
TI TDA4VM & TI J721EXCPXEVM & 2xA72 & 2000 & PPI & 162 & \textcolor{green!60!black}{Works} \\
TI AM69x & TI AM69 Starter Kit & 2x (4xA72) & 2000 & PPI & 163 & \textcolor{green!60!black}{Works} \\
Raspberry Pi 4 & Raspberry Pi 4 Model B & 4xA72 & 600 & - & - & CoreSight broken \\
RK3566 & Lubancat Zero N & 4xA55 & 816 & PPI-edge & 217 & \textcolor{green!60!black}{Works} \\
RK3568 & Youyeetoo YY3568 & 4xA55 & 816 & PPI-edge & 217 & \textcolor{green!60!black}{Works} \\
RK3588 & Firefly ITX-3588J & 4xA55+4xA76 & 1120+1200 & PPI-edge & 272+308 & \textcolor{green!60!black}{Works} \\
Orin & NVIDIA Jetson AGX Orin & 3x (4xA78) & 2000 & - & - & No CTIIRQ \\
\hline
\end{tabular}
\label{tab:boards}
\end{table*}

%% file: tab-ETM_correctness_table.tex
\begin{table*}[t]
\caption{\etmReg correctness evaluation for Cortex-A53/A72/A55/A76/A78. For each operation (\eg \emph{read}), the table shows the ratio between the memory bandwidth observed by PMU and ETM counters, versus the one observed by the \emph{bench} benchmark.}
\centering
\resizebox{\textwidth}{!}{
\begin{tabular}{lcrrrrrrrrrrrrrrrrrrrrrr}
\hline
Board & \multicolumn{1}{c}{Core} & \multicolumn{2}{c}{prefetch\_l1} & \multicolumn{2}{c}{prefetch\_l2} & \multicolumn{2}{c}{prefetch\_l3} & \multicolumn{2}{c}{read} & \multicolumn{2}{c}{read\_ldnp} & \multicolumn{2}{c}{write} & \multicolumn{2}{c}{write\_dczva} & \multicolumn{2}{c}{write\_stnp} & \multicolumn{2}{c}{modify} & \multicolumn{2}{c}{modify\_prefetch} & \multicolumn{2}{c}{modify\_stnp} \\
 &  & \multicolumn{1}{c}{PMU} & \multicolumn{1}{c}{ETM} & \multicolumn{1}{c}{PMU} & \multicolumn{1}{c}{ETM} & \multicolumn{1}{c}{PMU} & \multicolumn{1}{c}{ETM} & \multicolumn{1}{c}{PMU} & \multicolumn{1}{c}{ETM} & \multicolumn{1}{c}{PMU} & \multicolumn{1}{c}{ETM} & \multicolumn{1}{c}{PMU} & \multicolumn{1}{c}{ETM} & \multicolumn{1}{c}{PMU} & \multicolumn{1}{c}{ETM} & \multicolumn{1}{c}{PMU} & \multicolumn{1}{c}{ETM} & \multicolumn{1}{c}{PMU} & \multicolumn{1}{c}{ETM} & \multicolumn{1}{c}{PMU} & \multicolumn{1}{c}{ETM} & \multicolumn{1}{c}{PMU} & \multicolumn{1}{c}{ETM} \\
\hline
ideal &  & 1.00 & 1.00 & 1.00 & 1.00 & 1.00 & 1.00 & 1.00 & 1.00 & 1.00 & 1.00 & 1.00 & 1.00 & 1.00 & 1.00 & 1.00 & 1.00 & 2.00 & 2.00 & 2.00 & 2.00 & 2.00 & 2.00 \\
\hline
ZCU102 & A53 & 1.00 & 1.00 & 1.00 & 1.00 & - & - & 1.00 & 1.00 & 1.00 & 1.00 & 0.97 & 0.97 & 1.00 & 1.00 & 1.00 & 1.00 & 2.00 & 2.00 & 2.00 & 2.00 & 2.00 & 2.00 \\
TI AM69x & A72 & 1.76 & 1.74 & 1.76 & 1.73 & 1.76 & 1.73 & 1.76 & 1.74 & 1.76 & 1.74 & 1.00 & 1.00 & 1.00 & 1.00 & 1.00 & 1.00 & 1.76 & 1.75 & 1.76 & 1.74 & 1.00 & 1.00 \\
RK3588 & A55 & 1.93 & 1.91 & 1.94 & 1.90 & 0.00 & 0.00 & 1.92 & 1.81 & 1.92 & 1.80 & 1.00 & 1.00 & 1.00 & 1.00 & 1.00 & 1.00 & 1.94 & 1.93 & 1.94 & 1.91 & 1.93 & 1.88 \\
RK3588 & A76 & 0.30 & 0.29 & 0.27 & 0.27 & 0.06 & 0.06 & 2.00 & 1.90 & 1.21 & 1.19 & 1.00 & 1.00 & 1.00 & 1.00 & 1.00 & 1.00 & 2.00 & 1.94 & 2.00 & 1.93 & 1.00 & 1.00 \\
Orin & A78 & 0.04 & 0.04 & 0.01 & 0.01 & 0.01 & 0.01 & 2.00 & 2.00 & 2.00 & 2.00 & 2.00 & 2.00 & 2.00 & 2.00 & 2.00 & 2.00 & 2.00 & 2.00 & 2.00 & 2.00 & 2.00 & 2.00 \\
\hline
\end{tabular}}
\label{tab:correctness}
\vspace{-1em}
\end{table*}

%% file: 5-evaluation.tex
\section{Evaluation}
\label{sec:eval}


The evaluation addresses the following questions.

\textbf{Correctness:}
Is \etmReg correctly accounting for LLC refills/write-backs
and hitting the bandwidth regulation target.

\textbf{Boundaries:}
What are the 
resolution boundaries of \etmReg regulation,
and what are the most appropriate 
settings.

\textbf{Comparison:}
How does \etmReg perform in comparison to state-of-the-art approaches, namely \memguard and \mempol.


\subsection{Setup}

We evaluated the applicability of \etmReg across multiple SoCs and core types,
\ie Cortex-A53, A72, A55, A76 and A78.
Table~\ref{tab:boards} summarizes tested boards and their settings. 

To evaluate the
correctness and the boundaries of \etmReg,
we run micro-benchmarks on an RTOS,
which provides a low-noise environment and allows us to perform measurements close to the hardware without unwanted interference of unrelated tasks, like on a complex OS such as Linux.
We use the \emph{bench} memory bandwidth benchmark~\cite{bench} as load generator and to measure the behavior of different memory access types, \ie \emph{prefetch}, \emph{read}, \emph{write}, and \emph{modify}. 



We compare \etmReg with \memguard and \mempol using the same setup
that was also used in previous studies like~\cite{ZuepkeBCCM24, IzhbirdeevH0ZHC24}.
Specifically, we use a ZCU102 board featuring a Zynq UltraScale+ SoC with four Cortex-A53 cores.
Our test system runs Debian 12 with a Linux v6.1 vendor kernel from AMD/Xilinx.
We use the \emph{San Diego Vision Benchmark Suite} (\emph{SD-VBS})~\cite{venkata2009sd}
in the context of \emph{RT-Bench}~\cite{rt-bench}.
To test the isolation properties of different regulation techniques,
we use 
\emph{IsolBench}~\cite{isolbench} 
as an interference generator (\code{bandwidth} \emph{write} of 16~MiB size).%

\subsection{Correctness of ETM-based Budget Accounting}
\label{eval:correctness}


To verify the correctness of the proposed regulation mechanisms,
we need to first verify that the ETM observes the correct sum of the cache-related counters used for regulation on the specific cores, as the ETM can only aggregate the low-level cache signals into one \emph{OR-ed} input.
For this, we configure the ETM into a pure counter mode,
\ie we chain the two counters into a single 32-bit counter
and disable generation of interrupts,
and compare both ETM and PMU counters for various memory access operations
over a 32~MiB-sized memory region.
%
For ETM and PMCs, we use the counters listed in Sec.~\ref{sec:impl-core-settings}.
Memory access variants with \emph{ldnp} or \emph{stnp}
indicate that non-temporal loads and stores were used, which might bypass caching.
\emph{Write_dczva} zeros a full cache line with the Arm \emph{DC ZVA} instruction.


Table~\ref{tab:correctness} shows the results on five representative SoCs to cover each evaluated core type.
The table compares the ratio between the memory bandwidth observed by PMU and ETM counters, versus the one observed by the \emph{bench} benchmark.
A value of 1.0 for \emph{prefetch}, \emph{read}, and \emph{write} variants
resp.\@ 2.0 for \emph{modify} (as it contains both reads and writes)
indicates a perfect match between PMU resp.\@ ETM counters and the observed bandwidth, as shown in row \emph{ideal}.
Values $<1$ show that PMU or ETM counters undercount the memory bandwidth;
the regulator is too optimistic.
In contrast,
values of $>1$ show overcounting,
\ie the accounted value is pessimistic.

The Cortex-A53 (Zynq UltraScale+) offers
a good match of the selected PMU events for regulation.
The ETM counter also matches.
The Cortex-A53 core implements prefetches from L3 cache as NOP.

The Cortex-A72 (AM69x) overcounts
\emph{reads} and some \emph{\mbox{modify}} operations by 75\%
(leading to undercounting for \emph{modify}).
Here, the ETM counter misses a small percentage of the events observed by the PMC.

The Cortex-A55 (RK3588, $B_{A55}^\emph{moderate2}$)
expectedly overcounts \emph{reads}
and slightly undercounts \emph{modify} operations.
%
The ETM counter misses a small percentage of events.

On the Cortex-A76 (RK3588, $B_{A76}^\emph{moderate2}$),
\emph{prefetches} are not correctly accounted for,
and \emph{reads} are expectedly overcounted.
%
The ETM counter misses about 7\% of events.

For the Cortex-A78 (Nvidia Orin, $B_{A78}^\emph{pessimistic}$),
we see the same problem for \emph{prefetches} as on the Cortex-A76.
However, the measurement shows a good match of both PMCs and ETM accounting to the ideal baseline.

Overall,
the measured results are in line with previous results~\cite{PradhanOJHZZBC25}.
This shows that the accounting of PMU events by the ETM
is appropriate for 
memory bandwidth regulation.

A second important aspect for the regulation is the latency \emph{from} observing an event at the ETM \emph{to} throttling in the interrupt handler.
To measure this latency,
%
we program the ETM to trigger an interrupt
when it observes an unaligned load/store operation that crosses two cache lines,
%
and
we can observe that an interrupt becomes pending to the core by polling the \emph{ISR_EL1} register.
We then measure the time in CPU cycles 
from the unaligned memory access to the arrival of the interrupt at the core,
effectively bypassing any operating system overheads.
As shown in Table~\ref{tab:boards} (column \emph{IRQ latency}),
the latency is up to a quarter of a microsecond on all platforms.


\input{fig-error}

\input{fig-mempol}


%
%

\subsection{Regulation Behavior at Small Regulation Periods}
\label{eval:error}


We evaluate the regulation behavior at small regulation periods in Fig.~\ref{fig:error},
and explain two design decisions in \etmReg.
We construct two modified variants of \etmReg \memguardBased, 
\memguardBased-stop and \memguardBased-user, 
and compare the regulation behavior
for a replenishment period of 5~µs and small budget values.

%
%
\memguardBased (Fig.~\ref{fig:error} (left))
is shown as reference and regulates as expected
after a minimum bandwidth is achieved
(see also Sec.~\ref{eval:boundaries}).
For low bandwidth targets,
the regulation misses its target
by going over the two diagonal dashed lines
(top dashed line for \emph{prefetch}, \emph{read}, and \emph{write}; bottom dashed line for \emph{modify}---recall from Sec.~\ref{sec:impl-core-settings} that modify comprises both read and write parts and achieves only half of the bandwidth target, however the regulator observes the sum of the PMCs, as the purple dashed line shows),
but eventually converges for bandwidths $\geq$350~MB/s.
We discuss this effect in Sec.~\ref{eval:boundaries}.

\memguardBased-stop (Fig.~\ref{fig:error} (center)) \emph{stops accounting} PMU events if the budget is exceeded, like \memguard.
The regulation misses its target and shows a systematic \emph{regulation error}
that can be attributed to:
%
%
(i) the latency of CTIIRQ,
(ii) buffering of outstanding transactions in the memory subsystem,
and 
(iii) the fact that the operating system 
also performs memory accesses during interrupt handling.
We observe that 
\emph{read} and \emph{prefetch} exceed the target by up to eight cachelines
(the number of cachelines is shown on top and applies to the y-axis as well).
We attribute this error to latency effects,
as the Cortex-A53 core has limited out-of-order capabilities on loads.
For \emph{write} and \emph{modify}
the absolute error increases 
to about 20 cachelines due to the additional buffering of transactions.
We conclude that regulation at small scale must never stop accounting.

\memguardBased-user (Fig.~\ref{fig:error} (right)) accounts for PMU events in \emph{user space} only.
In our experiment,
which uses an RTOS with timer and regulation interrupts as kernel activities,
only \emph{modify} shows a regulation error,
which is likely caused by delays in accounting of buffered transactions,
as code and data is hot in the cache.
However, any kernel activity would contribute to regulation errors
if accounting is excluded for the kernel.

Note that these effects are mostly relevant for regulation at microsecond scale.
For larger periods,
these effects have less impact on regulation quality,
as their absolute error is small.


\input{fig-bench2a}
\input{fig-bench3}


\subsection{Boundaries and Limitations of the ETM² Regulation}
\label{eval:boundaries}


We show the limitations of \etmReg
when using budget values that are too small.
%
Fig.~\ref{fig:mempol-small} (left) shows the behavior of the \mempolBased~3:1 regulator
for a replenishment period of 5~µs, similar to Fig.~\ref{fig:error}. 
We observe that the regulation misses its target,
but eventually converges when the given budgets become larger.
This effect is caused by 
\emph{overflows of the PMC counter} in the ETM
(recall from 
Sec.~\ref{sec:design-mempol} that \etmReg accounts all the time).
%
%
Shifting more accounting capabilities toward the throttling state
mitigates the regulation error,
and the regulation hits the bandwidth target much earlier,
as in Fig.~\ref{fig:mempol-small} (middle) and (right)
for \mempolBased~2:2 resp.\@ \mempolBased~1:3.
Note that for such peak memory accesses at small budgets,
we can observe the same limitation in the \mempolBased regulator
and on all platform, as Sec.~\ref{sec:eval-all} shows.

However, \mempolBased~1:3's better regulation at small values comes at a cost,
as Fig.~\ref{fig:mempol-large} shows.
\mempolBased~1:3 (right) starts to oscillate once we reach peak bandwidth levels,
as the \mempolBased regulator triggers the throttling state
due to the missing reset of the ETM counter in the design,
see Sec.~\ref{sec:design-mempol}.
We therefore recommend using only the \mempolBased~3:1 and \mempolBased~2:2 variants
of the \mempolBased regulators,
and the former only for larger bandwidth values.
%


\subsection{Large-scale Evaluation}
\label{sec:eval-all}

Figs.~\ref{fig:bench-s32g} to~\ref{fig:bench-rk3588-a76} 
show the \memguardBased regulation on a wide range of platforms.
\ifversionCamera
Due to space constraints, results for 
further boards
are presented in the extended version of the paper~\cite{etm2-extended}.
\else
Results for further boards are presented the Appendix~\ref{sec:eval-all-appendix}.
\fi
%
As in Figs.~\ref{fig:error} to~\ref{fig:mempol-large},
%
the top graph shows the regulation behavior
for typical memory budgets on embedded platforms,
\ie less than 1000 MB/s,
for a small replenishment period of 5~µs (10~µs on RK3588).
The bottom graph shows the regulation behavior
for a larger replenishment period of 20~µs
and over a wider range of bandwidth targets
and allows for comparison of platform-specific factors.
As noted in Sec.~\ref{eval:boundaries},
these \memguardBased trends also apply to \mempolBased~3:1 and \mempolBased~2:2.
The bottom graphs in each figure primarily illustrate the limitations of each platform.
The top graphs show the minimum budget that is required for the
regulation to hit the target bandwidth.

On Cortex-A53-based SoCs
(Fig.~\ref{fig:bench-s32g}),
we observe similar trends at low memory bandwidth targets
as previously discussed 
in Sec.~\ref{eval:error} (Fig.~\ref{fig:error})
and Sec.~\ref{eval:boundaries} (Figs.~\ref{fig:mempol-small} and~\ref{fig:mempol-large}).
%

On Cortex-A72-based SoCs
(Figs.~\ref{fig:bench-lx2160a} and~\ref{fig:bench-am69}),
the trend is similar.
In general,
the Cortex-A72 cores can sustain a higher memory bandwidth,
as the platforms have more powerful memory controllers.
The regulation can keep the bandwidth target
for sufficiently large budgets and replenishment periods.

\input{fig-slowdown_sdvbs_1pmc}


On Cortex-A55-based SoCs
(Figs.~\ref{fig:bench-rk3568} and~\ref{fig:bench-rk3588-a55}),
we see similar trends.
However,
the RK3588 has a much more powerful memory controller,
so the regulation stabilizes later.
On the same SoC,
we can see that the memory subsystem of the Cortex-A76 is much more
powerful than the A55 (Fig.~\ref{fig:bench-rk3588-a76}).
Note that the top graphs in Figs.~\ref{fig:bench-rk3588-a55} and~\ref{fig:bench-rk3588-a76}
show the regulation at 10~µs; otherwise, the interesting range would be out of range.
The regulation follows the trends discussed in Table~\ref{tab:correctness},
and the regulation slightly undercounts for both Cortex-A55 and A76.

We were not able to evaluate \etmReg on Cortex-A78,
as our platform, the Nvidia Orin,
does not route CTIIRQ to the GIC.




\subsection{Comparison of ETM², MemGuard, and MemPol}
\label{eval:cmp}

We compare \etmReg with \memguard and \mempol using
\sdvbshort
on the Zynq UltraScale+ ZCU102
and on the RK3588.
For MemGuard and MemPol, we use the reference implementations provided by the authors.
MemGuard is implemented as a kernel module~\cite{YYPCS:13, YYPCS:16}.
The MemPol regulator runs on the first Cortex-R5 real-time core
in its default settings with a 6.25~µs polling interval and a sliding window size of 8 for regulation over a period of 50~µs~\cite{ZuepkeBCCM24}.

We use two experimental settings: the first
(similar to
~\cite{ZuepkeBCCM24})
uses a single PMC to track \emph{reads} and is the baseline for comparing
with \memguard (Fig.~\ref{fig:sdvbs-slowdown-1pmc}).
The primary goal is to show that the best variants of each
regulation strategy achieve comparable results.
The second setup (Fig.~\ref{fig:sdvbs-a55-a76}) uses different core-specific
regulation settings (see Sec.~\ref{sec:impl-core-settings}) to illustrate
the effects of pessimistic and moderate regulation strategies.
%
In both experiments, we run \sdvbshort benchmarks with \vga image size under regulation at different bandwidth settings.%
\ifversionCamera
\footnote{For space reason, Figs.~\ref{fig:sdvbs-slowdown-1pmc} and \ref{fig:sdvbs-a55-a76} only show the most relevant trends. Results for \emph{sift} and \emph{tracking} benchmarks are available in~\cite{etm2-extended}.}
\else
\footnote{Figs.~\ref{fig:sdvbs-slowdown-1pmc} and \ref{fig:sdvbs-a55-a76} only show the most relevant trends. Results for \emph{sift} and \emph{tracking} benchmarks are available in Appendix~\ref{eval:cmp-appendix}.}
\fi
The experiments distribute fractions of the,
empirically determined,
bandwidth available under worst case conditions (1000~MB/s on the ZCU102~\cite{ZuepkeBCCM24}, 2000~MB/s on the RK3588~\cite{PradhanOJHZZBC25})
between benchmarks and interference tasks.
The bandwidth targets for the benchmarks are 
20\%, 30\%, and 40\%.
%
The interference cores use and share the remaining bandwidth,
\ie 80\%, 70\%, and 60\%.
%
The first row shows the results for execution of \sdvbshort benchmarks in isolation.
The middle and bottom rows show the results with increasing cores 
running IsolBench~\cite{isolbench} as load generator and causing memory interference.
The graphs show the normalized \emph{slowdown} of execution times
compared to unregulated execution without interference (thin red line).
The colored bars indicate the average over 10 runs.
Small vertical black lines show min/max. 


Fig.~\ref{fig:sdvbs-slowdown-1pmc} shows the results for regulation based on cache refills.
Here, the regulation provided by \etmReg
%
nearly matches previously reported results for \memguard and \mempol~\cite{ZuepkeBCCM24}.
For \memguard, we observe that for small regulation periods like 50~µs and 20~µs, the overhead increases and often exceeds the other regulators.
\etmReg and \mempol show slightly higher slowdown on memory intensive benchmarks like \emph{disparity} and \emph{mser} than \memguard. 
%
Note that 
\memguard only regulates user space code
and does not account for kernel activities like system calls or interrupt handling,
see Sec.~\ref{eval:error}.
%
The memory intensive benchmarks also expose that \etmReg has slightly higher overhead than \mempol.
We attribute this to \etmReg's interrupt-based throttling,
but we have not further investigated this effect nor the timing differences to \memguard.
%
%
Otherwise, all regulators behave similarly on compute intensive benchmarks
like \emph{stitch} 
\ifversionCamera
(see also~\cite{etm2-extended}).
\else
(see also Appendix~\ref{eval:cmp-appendix})
\fi
%
Differences between \etmReg's \memguardBased and \mempolBased regulations are minimal,
with \mempolBased 50~µs being slightly faster.
All regulators show also a similar behavior in providing isolation from interference,
as the comparison between the top row (isolation, no interference) and the bottom rows (interference from IsolBench) shows.
Also, the regulation behavior of \etmReg nearly matches \mempol's.
The results provide empirical evidence that \emph{OR}-ing is not a problem for both real-world applications and micro-benchmarks.


\input{fig-slowdown-a55-a76-rk3588}


Fig.~\ref{fig:sdvbs-a55-a76} shows the results
on newer Cortex-A55 and A76 cores (RK3588)
for regulation with the settings from Sec.~\ref{sec:impl-core-settings}.
Here, in addition to the \emph{pessimistic} regulation settings, 
both \etmReg and MemPol can use the less pessimistic \emph{moderate} ones.%
\footnote{
Due to space constraints,
we compare only the best configurations of the regulators
and limit interference to 1x~IsolBench running on 1x~Cortex-A55 and 4x~IsolBench running on 2×~Cortex-A55 and 2×~Cortex-A76.
}
%
As expected, more accurate accounting is beneficial over
\emph{pessimistic} regulation models.
Similarly to Fig.~\ref{fig:sdvbs-slowdown-1pmc}, under pessimistic settings,
MemGuard behaves comparably or slightly better than the other regulators,
but, when available on the target platform, moderate settings proves much more
effective in achieving lower slowdowns.
We also note that on newer cores, the impact of enforcing a reduced bandwidth is much more pronounced that on older cores. This can be observed on, \eg \emph{disparity} on the Cortex-A76, where the slowdown reaches up to factor 80.
%
%
%
Lastly, we can observe that memory bandwidth regulation alone does not protect from other types of interference, \eg in the shared LLC, as the differences between the top row and the bottom rows show.

\subsection{Discussion}
\label{sec:discussion}

Overall, \etmReg behaves as intended, delivering regulation
performance that closely matches \memguard and \mempol.
For small regulation periods, \etmReg shows less overhead than \memguard,
as no timer interrupts for periodic replenishment are needed.
Like \mempol, \etmReg depends on the CoreSight debugging infrastructure.
However, \etmReg does not require a dedicated core for regulation purposes,
as the ETM state machine implements the regulation.

Although the use of \emph{OR}-ed inputs can lead to undercounting,
our experiments indicate that this effect is minimal in practice
and does not affect regulation accuracy.
The effect of overflowing the PMU event counter in the ETM regulation,
leading to errors in the regulation,
can be mitigated by using larger budgets,
larger periods,
or by using the \mempolBased regulator instead of \memguardBased.
In any case, our experiments provide baseline thresholds
for safe operation of the regulation on each platform.

The leeway of the token-bucket regulation in the \memguardBased regulator 
to compensate for memory bursts is quite limited,
but sufficient to extend the regulator's period over a longer time span
than what is possible with a single 16-bit cycle counter.
We would have wished for more states in the ETM sequencer or additional counters to cope with this limitation.

Another technical challenge for \etmReg is the lack of precise events to monitor cores' LLC activity on modern DSU-based Arm cores like Cortex-A55 and A76.
The bandwidth regulators proposed in~\cite{PradhanOJHZZBC25} were instrumental for the implementation and evaluation on the large scale of SoCs, but are limited in the precision \wrt monitoring of cores' memory activity.
%
Table~\ref{tab:correctness} shows that \etmReg, similar to \mempol,
can use the \emph{optimistic} bandwidth regulation schemes,
while \memguard must use the \emph{pessimistic} ones~\cite{PradhanOJHZZBC25}.
%
Overall, \mempol allows for the greatest flexibility in combining multiple PMCs,
as it allows arbitrary factors for each PMC.


As the comparison to \memguard shows,
long regulation periods compensate for memory bursts better than short regulation periods.
This is a known, positive effect, especially
when bandwidth regulation is only needed to balance memory accesses between cores
at the software-level and with a resolution that matches the tasks' execution times.
Instead, the fine-grained regulation in \etmReg and \mempol allows co-regulation
at a much lower scale and at the hardware level of other busmasters in the system,
similar to QoS features of Arm interconnects~\cite{arm-qos400}.
%
Ultimately, the selection of the regulator depends on the application.

%% file: fig-error.tex
\begin{figure*}[!t]
\centering
	\includegraphics[width=0.665\columnwidth]{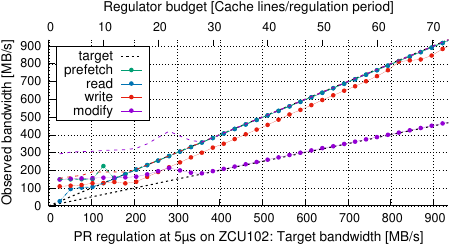}%
\hspace{0.5\columnsep}%
    \includegraphics[width=0.665\columnwidth]{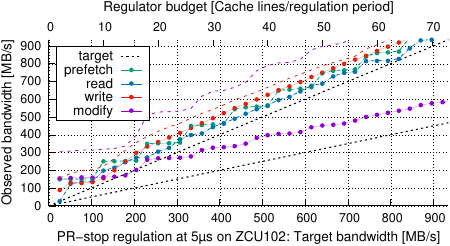}%
\hspace{0.5\columnsep}%
    \includegraphics[width=0.665\columnwidth]{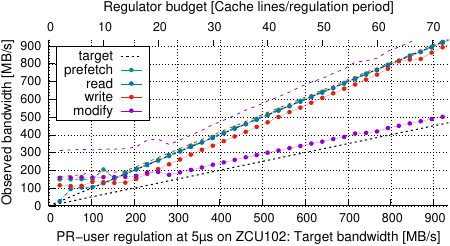}%
\caption{
    \etmReg's \memguardBased regulator variants on ZCU102 show different types of \emph{regulation errors} relevant at small regulation periods.
    PR (left, reference) regulates as expected at $\geq$350~MB/s. 
    PR-stop (center) \emph{stops accounting} of PMU events if the core's budget is exceeded, but ongoing memory transactions finish later. 
    %
    PR-user (right) accounts PMU events in \emph{user space} only.
    Cache-line eviction by \emph{e.g.\@} interrupt handlers is not accounted for.
    %
    %
    In all three graphs, colored dashed lines show observed PMC values.
}%
\label{fig:error}%
\vspace{-0.5em}
\end{figure*}%

%% file: fig-mempol.tex
\begin{figure*}[!t]
\centering
	\includegraphics[width=0.665\columnwidth]{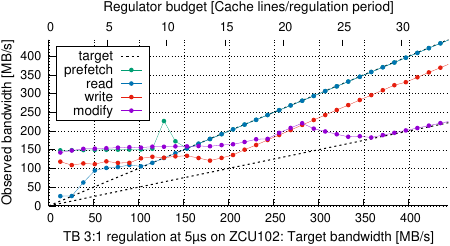}%
\hspace{0.5\columnsep}%
    \includegraphics[width=0.665\columnwidth]{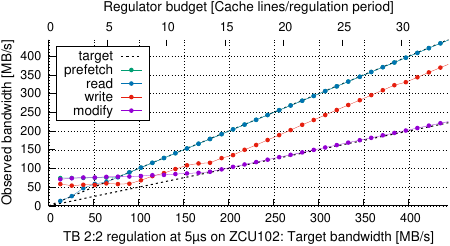}%
\hspace{0.5\columnsep}%
    \includegraphics[width=0.665\columnwidth]{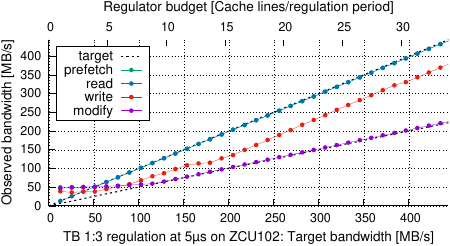}%
\caption{
    \etmReg's \mempolBased regulators on ZCU102 for small regulation values (bandwidth targets).
    %
    %
    \mempolBased~3:1 (left) shows problems of PMU counter overflows for small budget values as its PMU counter overflows multiple times.
    \mempolBased~2:2 (middle) is more robust to counter overflows and allows smaller regulation values.
    \mempolBased~1:3 (right) improves further.
}%
\label{fig:mempol-small}%
\vspace{-0.5em}
\end{figure*}%

\begin{figure*}[!t]
\centering
	\includegraphics[width=0.665\columnwidth]{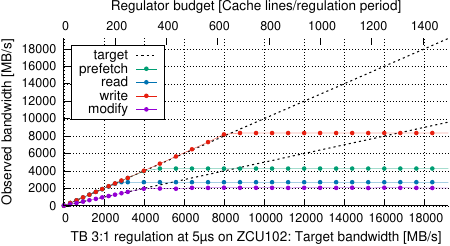}%
\hspace{0.5\columnsep}%
	\includegraphics[width=0.665\columnwidth]{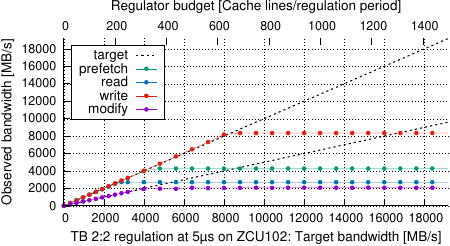}%
\hspace{0.5\columnsep}%
	\includegraphics[width=0.665\columnwidth]{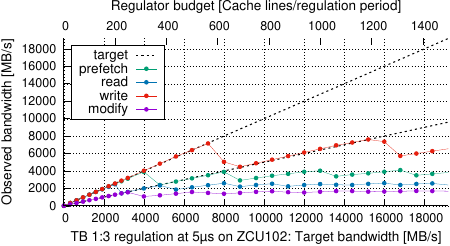}%
\caption{
    \etmReg's \mempolBased regulators on ZCU102 for large regulation values (bandwidth targets).
    %
    %
    \mempolBased~3:1 (left) and \mempolBased~2:2 (middle) regulate correctly at large regulation values.
    \mempolBased~1:3 (right) oscillates at peak bandwidth levels.
}%
\label{fig:mempol-large}%
\vspace{-0.5em}%
\end{figure*}%

%% file: fig-bench2a.tex
\begin{figure*}[!t]%
\begin{minipage}[t]{0.665\columnwidth}%
    \begin{tabular}{@{}c@{}}%
    \includegraphics[width=\linewidth]{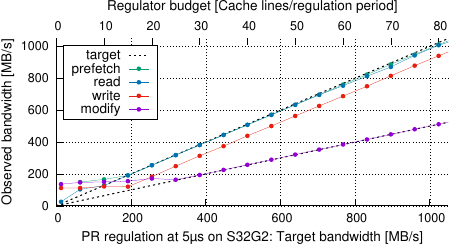}%
    \vspace{+0.2em}%
    \\%
    \includegraphics[width=\linewidth]{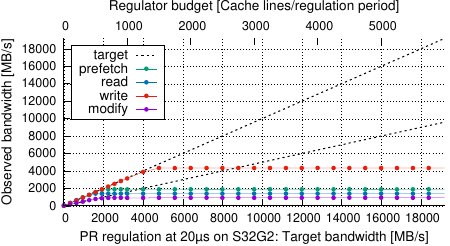}%
    \end{tabular}%
\caption{
    S32G2 with 4x Cortex-A53.
}%
\label{fig:bench-s32g}%
\end{minipage}%
%
\hfill%
\begin{minipage}[t]{0.665\columnwidth}%
    \begin{tabular}{@{}c@{}}%
    \includegraphics[width=\linewidth]{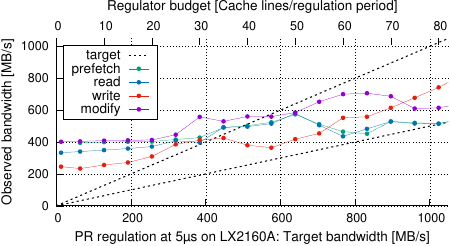}%
    \vspace{+0.2em}%
    \\%
    \includegraphics[width=\linewidth]{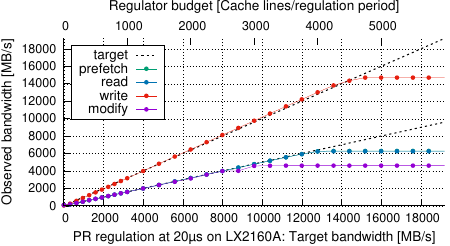}%
    \end{tabular}%
\caption{
    %
    %
    LX2160A with 16x Cortex-A72.
}%
\label{fig:bench-lx2160a}%
\end{minipage}%
%
\hfill%
%
\begin{minipage}[t]{0.665\columnwidth}%
    \begin{tabular}{@{}c@{}}%
	\includegraphics[width=\linewidth]{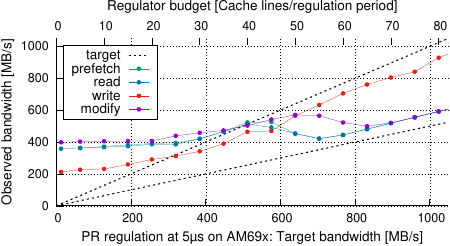}%
    \vspace{+0.2em}%
    \\%
	\includegraphics[width=\linewidth]{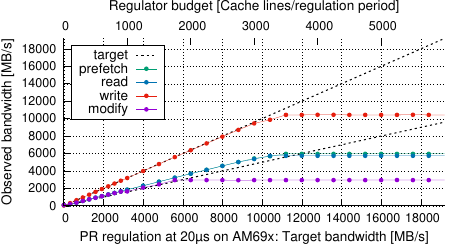}%
    \end{tabular}%
\caption{
    AM69x with 8x Cortex-A72.
}%
\label{fig:bench-am69}%
\end{minipage}%
\vspace{-0.5em}%
\end{figure*}%

%% file: fig-bench3.tex
%
%
%
\begin{figure*}[!t]%
\begin{minipage}[t]{0.665\columnwidth}%
    \begin{tabular}{@{}c@{}}%
	\includegraphics[width=\linewidth]{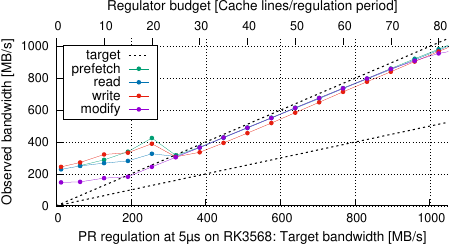}%
    \vspace{+0.2em}%
    \\%
	\includegraphics[width=\linewidth]{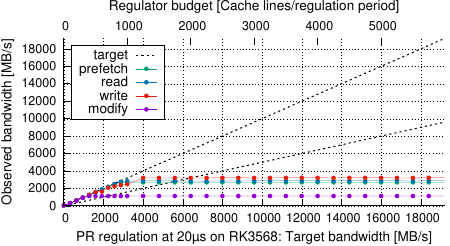}%
    \end{tabular}%
\caption{
    RK3568 with 4x Cortex-A55.
}%
\label{fig:bench-rk3568}%
\end{minipage}%
%
\hfill%
%
\begin{minipage}[t]{0.665\columnwidth}%
    \begin{tabular}{@{}c@{}}%
	\includegraphics[width=\linewidth]{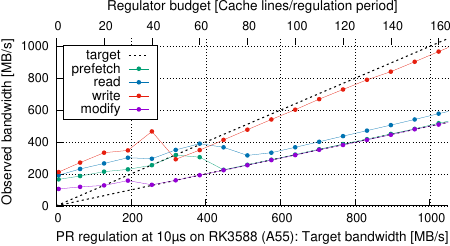}%
    \vspace{+0.2em}%
    \\%
	\includegraphics[width=\linewidth]{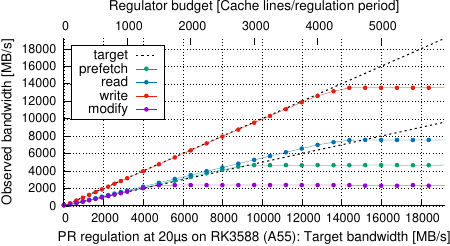}%
    \end{tabular}%
\caption{
    %
    %
    RK3588 with 4x Cortex-A55.
}%
\label{fig:bench-rk3588-a55}%
\end{minipage}%
%
\hfill%
%
\begin{minipage}[t]{0.665\columnwidth}%
    \begin{tabular}{@{}c@{}}%
	\includegraphics[width=\linewidth]{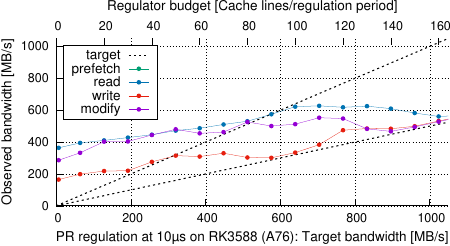}%
    \vspace{+0.2em}%
    \\%
	\includegraphics[width=\linewidth]{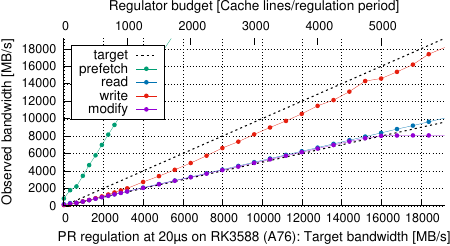}%
    \end{tabular}%
\caption{
    %
    %
    RK3588 with 4x Cortex-A76.
}%
\label{fig:bench-rk3588-a76}%
\end{minipage}%
\vspace{-1em}%
\end{figure*}%

%% file: fig-slowdown_sdvbs_1pmc.tex
\begin{figure*}[htbp]
    \begin{subfigure}{\textwidth}
        \centering
        \includegraphics[width=0.9\textwidth]{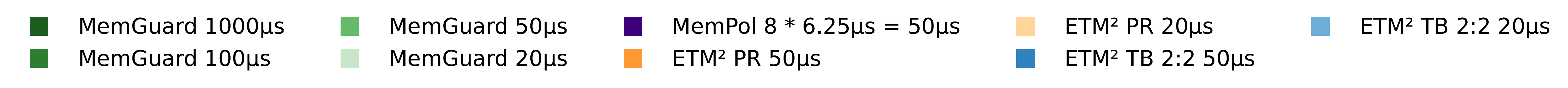}
    \end{subfigure}
    \centering
    \begin{subfigure}{\textwidth}
        \centering
        \includegraphics[width=\textwidth]{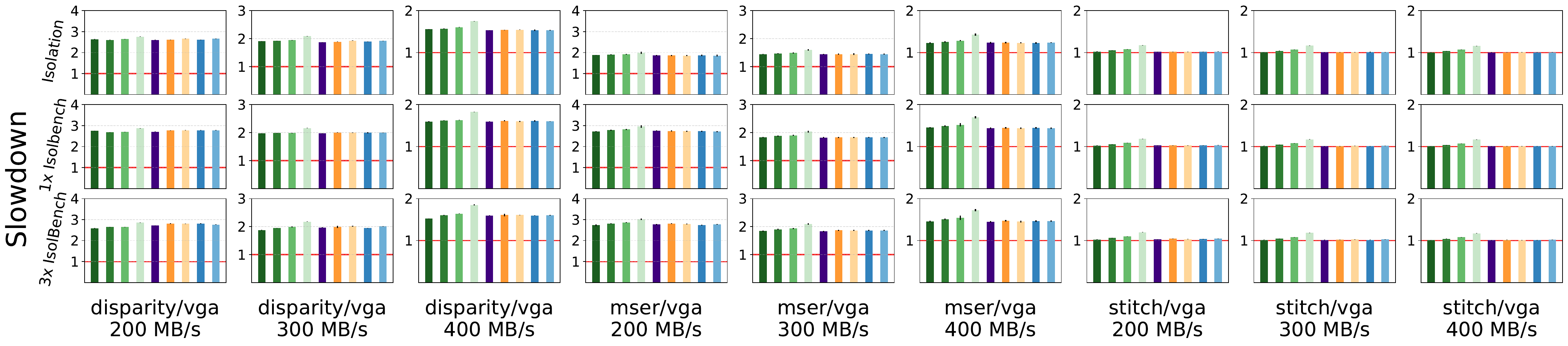}
        \captionsetup{justification=centering}
    \end{subfigure}

    \caption{Slowdown of SD-VBS benchmarks at different bandwidth targets (columns) 
    and increasing interference (rows)
    regulated by MemGuard, MemPol, and \etmReg 
    using a single PMC (\emph{L2D\_CACHE\_REFILL}) on ZCU102.
    See Sec.~\ref{eval:cmp} for details.}
    \label{fig:sdvbs-slowdown-1pmc}
\vspace{-1em}
\end{figure*}

%% file: fig-slowdown-a55-a76-rk3588.tex

\begin{figure*}[htbp]
    \begin{subfigure}{\textwidth}
        \centering
        \includegraphics[width=0.9\textwidth]{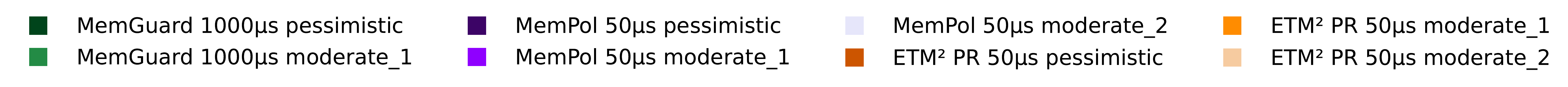}
    \end{subfigure}
    \centering
    \begin{subfigure}{\textwidth}
        \centering
        \subcaptionbox{Cortex-A55 (RK3588)} {
        \includegraphics[width=\textwidth]{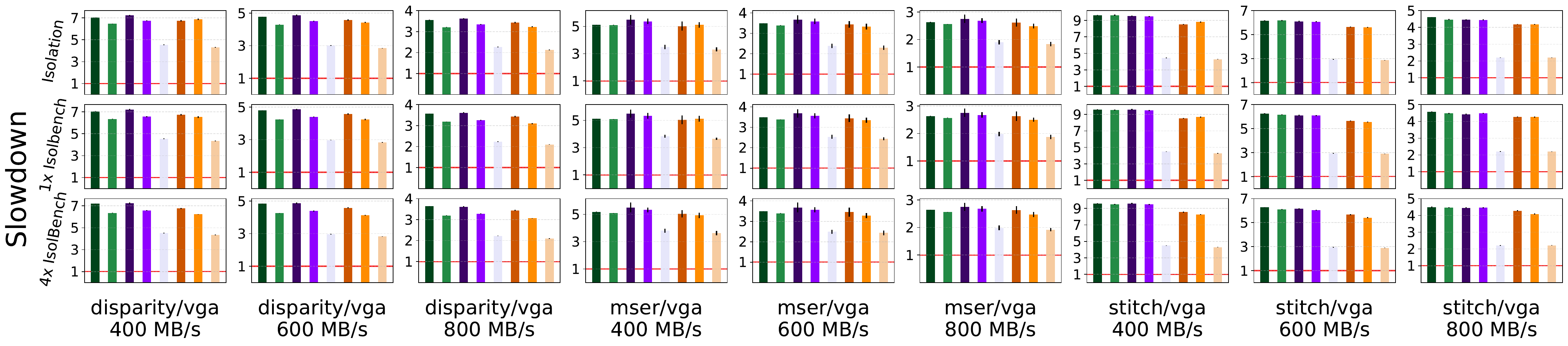}
        }
    \end{subfigure}
    \begin{subfigure}{\textwidth}
        \centering
        \subcaptionbox{Cortex-A76 (RK3588)} {
        \includegraphics[width=\textwidth]{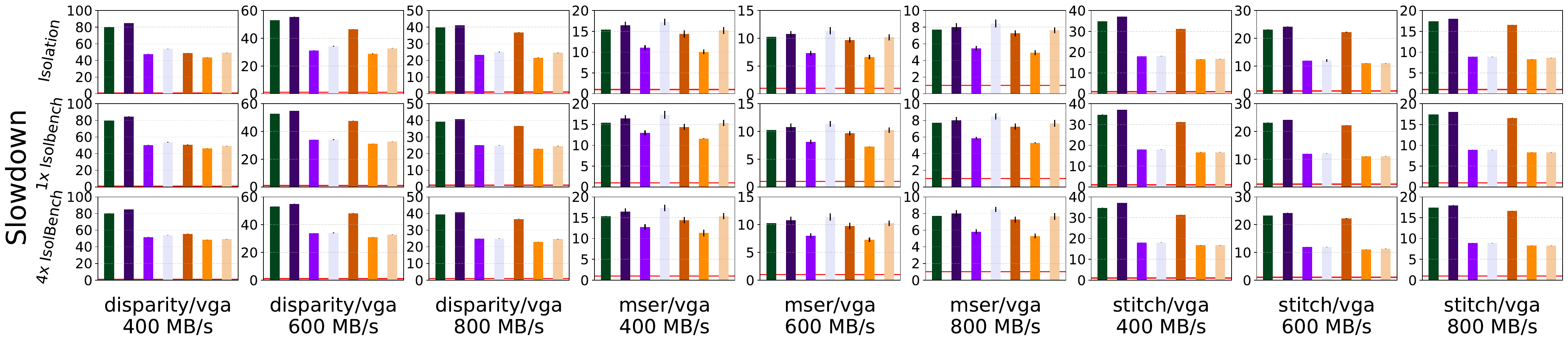}
        }
    \end{subfigure}
    
    \caption{Slowdown of SD-VBS benchmarks at different bandwidth targets (columns)
    and increasing interference (rows)
    regulated by MemGuard, MemPol, and \etmReg using the regulation settings from
    Sec.~\ref{sec:impl-core-settings} on the RK3588. See Sec.~\ref{eval:cmp} for details.}
    \label{fig:sdvbs-a55-a76}
\vspace{-1em}
\end{figure*}

%% file: 6-conclusion.tex
\section{Conclusion}
\label{sec:conclusion}

In this paper, we have shown that the Arm's CoreSight ETM
can be repurposed to implement an effective hardware-assisted
memory-bandwidth regulator for real-time multicore systems.
Our \etmReg design achieves microsecond-scale regulation intervals
and fast reaction times without relying on PMC polling
or dedicated external control cores.
Our novel approach complements the strengths of prior regulators:
it offers the fine-grained precision and multi-dimensional monitoring
of \mempol, while preserving the responsiveness of interrupt-driven approaches like \memguard.
%
We evaluated the correctness and capabilities of \etmReg on a broad
range of 64-bit Arm SoCs, demonstrating its wide portability.
Furthermore, we compared with previous results from \memguard and \mempol
in a similar evaluation context based on ZCU102 
and \sdvbshort.

Overall, \etmReg can scale efficiently and
enables regulation capabilities that were previously
infeasible with \memguard, without requiring dedicated cores as in \mempol.

Future work will focus on evaluating similar hardware support capabilities
on new generations of Arm v9 processors.

%% file: 7-appendix.tex
%
%
\clearpage
\afterpage{\clearpage}
\onecolumn

\appendix
\label{sec:appendix}

\subsection{Large-scale Evaluation (Additional Boards)}
\label{sec:eval-all-appendix}

We extend the large-scale evaluation presented in Sec.~\ref{sec:eval-all} 
for the other boards listed in Table~\ref{tab:boards}.
%
As in Figs.~\ref{fig:bench-s32g} to~\ref{fig:bench-rk3588-a76},
the top graphs show the regulation behavior
for typical memory budgets on embedded platforms,
\ie less than 1000 MB/s,
for a small replenishment period of 5~µs.
The bottom graph shows the regulation behavior
for a larger replenishment period of 20~µs
and over a wider range of bandwidth targets.

On Cortex-A53-based SoCs
(Figs.~\ref{fig:bench-zcu102} to~\ref{fig:bench-am67}),
we observe similar trends as
in Figs.~\ref{fig:error} to~\ref{fig:mempol-large} for the AMD/Xilinx Zynq UltraScale+
and Fig.~\ref{fig:bench-s32g} for the S32G.
Differences are in the available memory bandwidth,
but not the regulation behavior.

On Cortex-A72-based SoCs
(Fig.~\ref{fig:bench-tda4vm}),
the TDA4VM (two cores) compares to the related AM69x (eight cores, Fig.~\ref{fig:bench-am69}).

On Cortex-A55-based SoCs
(Fig.~\ref{fig:bench-rk3566}),
the behavior of the RK3566 compares to the RK3568 (Fig.~\ref{fig:bench-rk3568})
from the same SoC family,
but the memory controller differs.

\subsection{Comparison of ETM², MemGuard, and MemPol (Results for Additional Benchmarks)}
\label{eval:cmp-appendix}

We provide additional results
for the comparison between
\etmReg, \memguard, and \mempol
presented in Sec.~\ref{eval:cmp}.

Similar to
Fig.~\ref{fig:sdvbs-slowdown-1pmc},
Fig.~\ref{fig:sdvbs-slowdown-1pmc-appendix}
includes additional results for \sdvbshort benchmarks \emph{sift} and \emph{tracking}
on the ZCU102.

Fig.~\ref{fig:sdvbs-a55-a76-appendix}
extends Fig.~\ref{fig:sdvbs-a55-a76}
with additional results for \emph{sift} and \emph{tracking}
on the RK3588.


\input{fig-bench4-appendix}
\input{fig-bench5-appendix}
\input{fig-slowdown_sdvbs_1pmc-appendix}
\input{fig-slowdown-a55-a76-rk3588-appendix}

%% file: fig-bench4-appendix.tex
\begin{figure*}[!th]%
\begin{minipage}[t]{0.33\columnwidth}%
    \begin{tabular}{@{}c@{}}%
    \includegraphics[width=\linewidth]{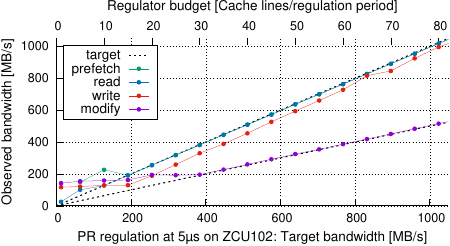}%
    \vspace{+0.2em}%
    \\%
    \includegraphics[width=\linewidth]{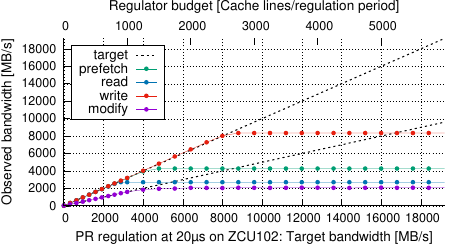}%
    \end{tabular}%
\caption{
    Zynq UltraScale+ with 4x A53.
}%
\label{fig:bench-zcu102}%
\end{minipage}%
%
\hfill%
%
\begin{minipage}[t]{0.33\columnwidth}%
    \begin{tabular}{@{}c@{}}%
	\includegraphics[width=\linewidth]{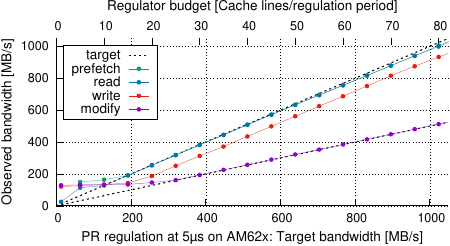}%
    \vspace{+0.2em}%
    \\%
	\includegraphics[width=\linewidth]{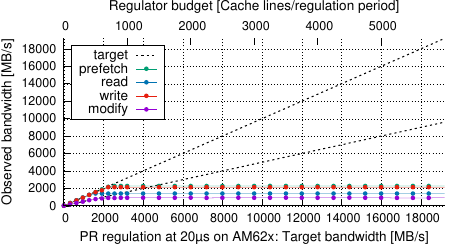}%
    \end{tabular}%
\caption{
    AM62x with 4x Cortex-A53.
}%
\label{fig:bench-am62}%
\end{minipage}%
%
\hfill%
%
\begin{minipage}[t]{0.33\columnwidth}%
    \begin{tabular}{@{}c@{}}%
    \includegraphics[width=\linewidth]{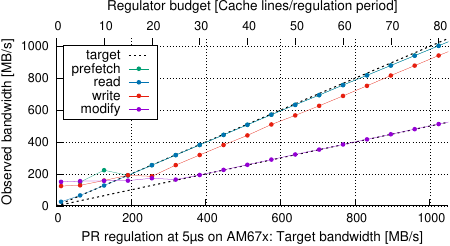}%
    \vspace{+0.2em}%
    \\%
    \includegraphics[width=\linewidth]{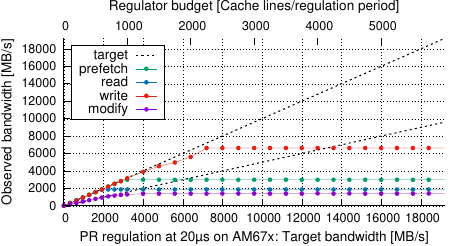}%
    \end{tabular}%
\caption{
    AM67x with 4x Cortex-A53.
}%
\label{fig:bench-am67}%
\end{minipage}%
\vspace{-1em}%
\end{figure*}%

%% file: fig-bench5-appendix.tex
\begin{figure*}[!th]%
\begin{minipage}[t]{0.33\columnwidth}%
    \begin{tabular}{@{}c@{}}%
    \includegraphics[width=\linewidth]{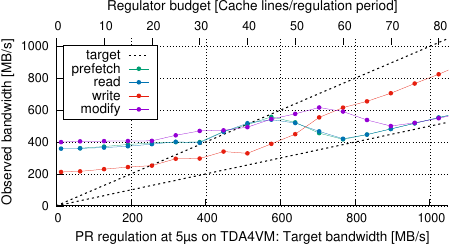}%
    \vspace{+0.2em}%
    \\%
    \includegraphics[width=\linewidth]{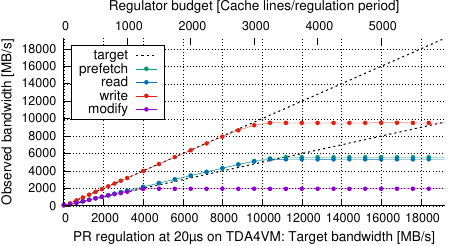}%
    \end{tabular}%
\caption{
    TDA4VM with 2x Cortex-A72.
}%
\label{fig:bench-tda4vm}%
\end{minipage}%
%
\hfill%
%
\begin{minipage}[t]{0.33\columnwidth}%
    \begin{tabular}{@{}c@{}}%
    \includegraphics[width=\linewidth]{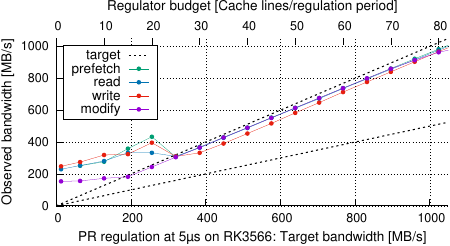}%
    \vspace{+0.2em}%
    \\%
    \includegraphics[width=\linewidth]{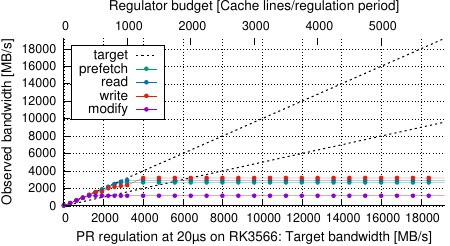}%
    \end{tabular}%
\caption{
    RK3566 with 4x Cortex-A55.
}%
\label{fig:bench-rk3566}%
\end{minipage}%
%
\hfill%
%
\begin{minipage}[t]{0.33\columnwidth}%
\mbox{} 
\end{minipage}%
\vspace{-1em}%
\end{figure*}%

%% file: fig-slowdown_sdvbs_1pmc-appendix.tex
\begin{figure*}[htbp]
    \begin{subfigure}{\textwidth}
        \centering
        \includegraphics[width=\textwidth]{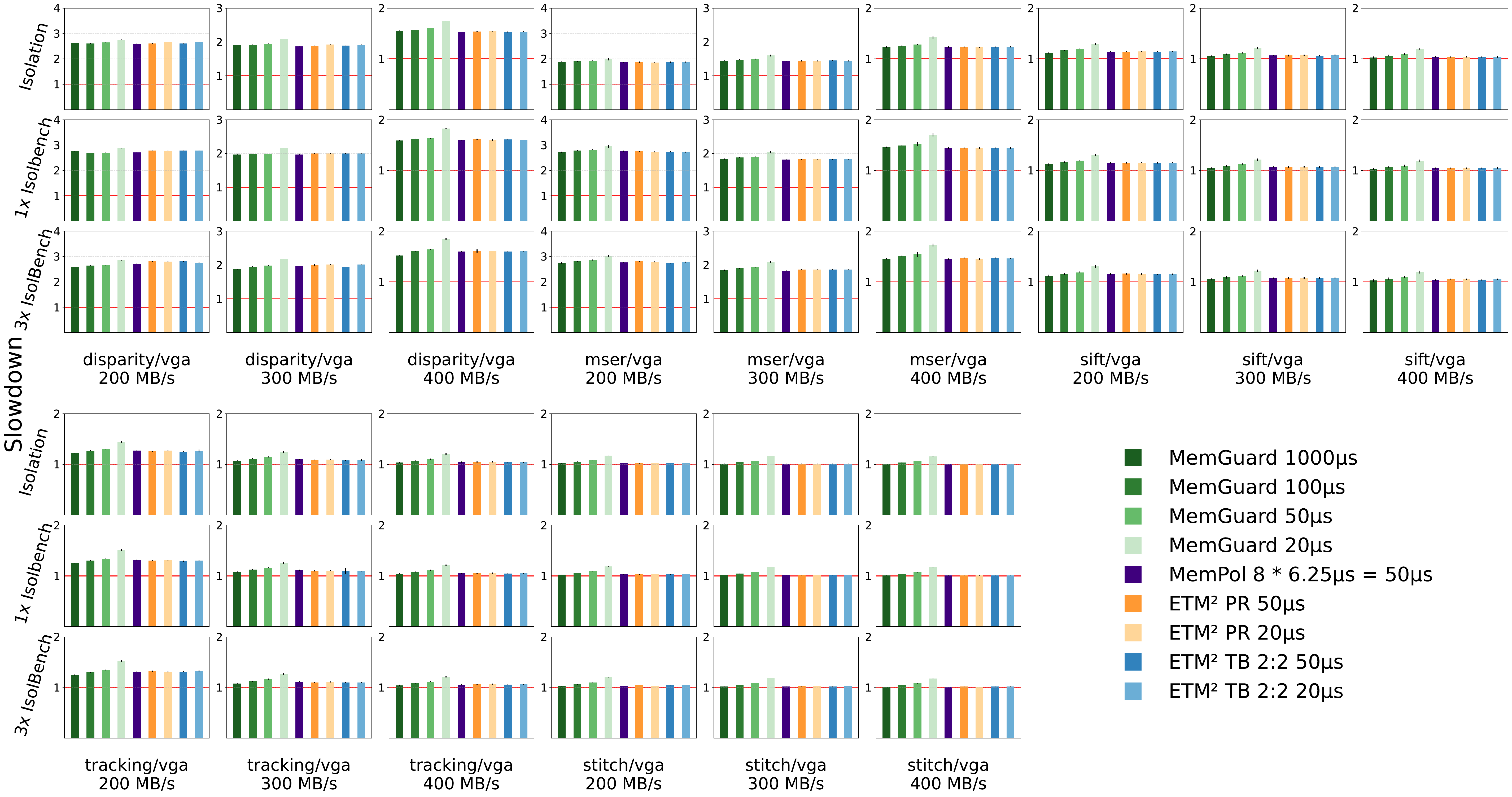}
        \captionsetup{justification=centering}
    \end{subfigure}
    \caption{Slowdown of SD-VBS benchmarks at different bandwidth targets (columns) 
    and increasing interference (rows)
    regulated by MemGuard, MemPol, and \etmReg 
    using a single PMC (\emph{L2D\_CACHE\_REFILL}) on ZCU102.
    See Sec.~\ref{eval:cmp} for details.
    }
    \label{fig:sdvbs-slowdown-1pmc-appendix}
\vspace{-1em}
\end{figure*}

%% file: fig-slowdown-a55-a76-rk3588-appendix.tex
\begin{figure*}[htbp]
    \begin{subfigure}{\textwidth}
        \centering
        \subcaptionbox{Cortex-A55 (RK3588)} {
        \includegraphics[width=\textwidth]{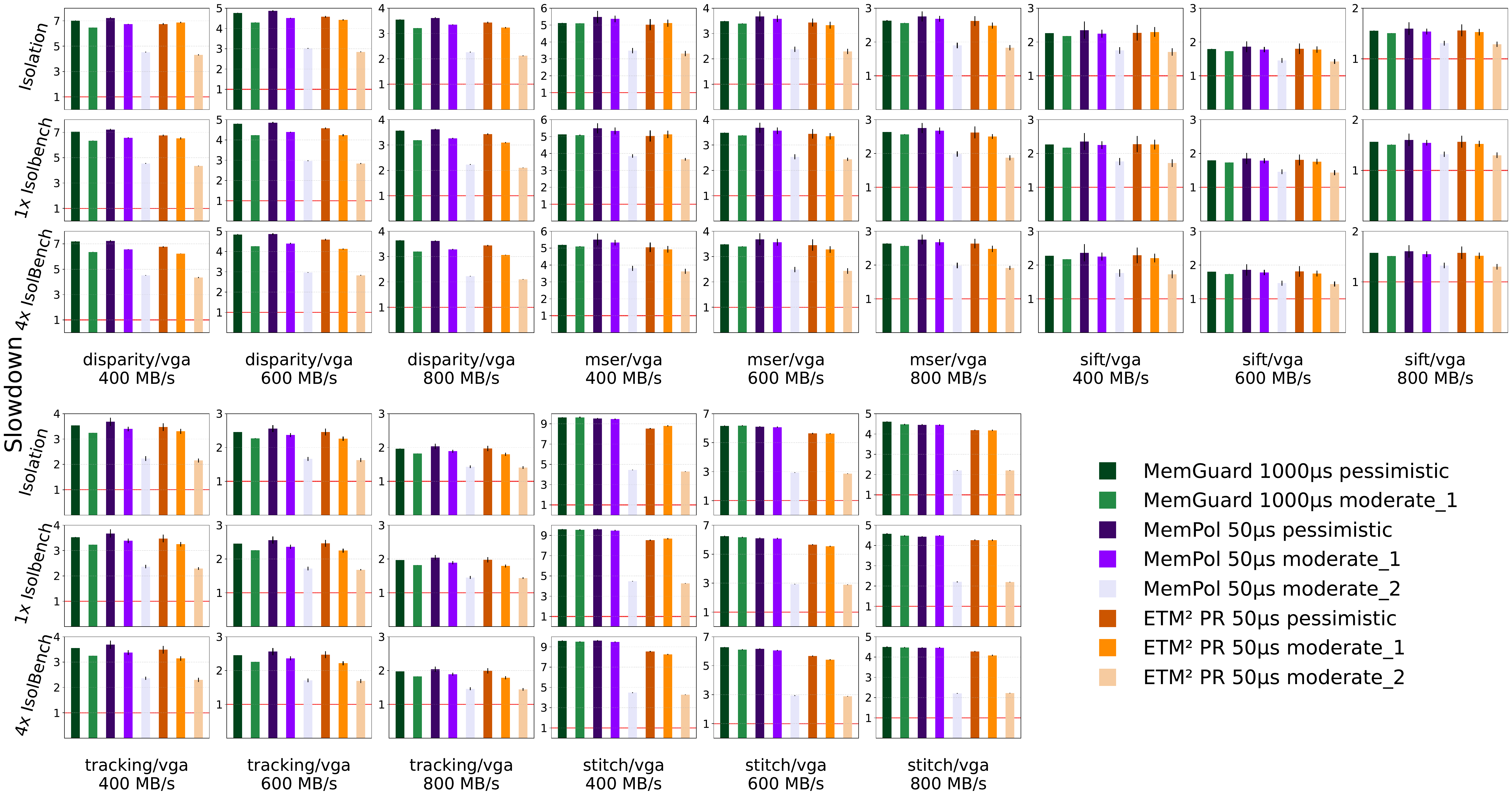}
        }
    \end{subfigure}
    \begin{subfigure}{\textwidth}
        \centering
        \subcaptionbox{Cortex-A76 (RK3588)} {
        \includegraphics[width=\textwidth]{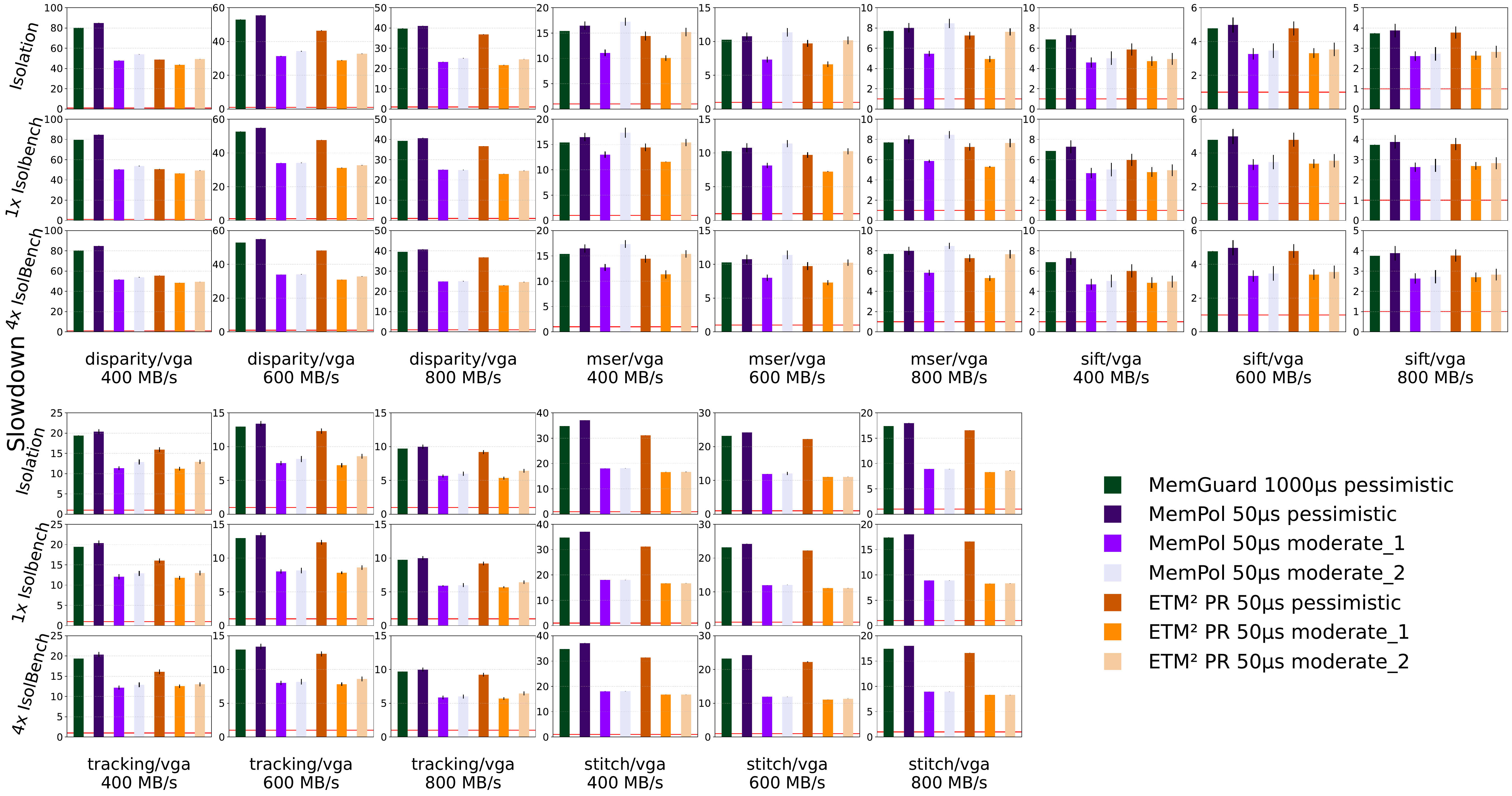}
        }
    \end{subfigure}
    
    \caption{Slowdown of SD-VBS benchmarks at different bandwidth targets (columns)
    and increasing interference (rows)
    regulated by MemGuard, MemPol, and \etmReg using the regulation settings from
    Sec.~\ref{sec:impl-core-settings} on the RK3588. See Sec.~\ref{eval:cmp} for details.
    }
    \label{fig:sdvbs-a55-a76-appendix}
\vspace{-1em}
\end{figure*}